\documentclass[12pt]{JHEP3}
\newcommand{\intd}{\mathrm{d}}
\newcommand{\ex}{\mathrm{e}}

\usepackage{amsmath,epsfig}
\usepackage{amssymb,amsfonts}
\usepackage{latexsym}
\usepackage{slashed}
\usepackage{empheq}
\numberwithin{equation}{section}
\usepackage{dsfont}
\usepackage{cancel}
\usepackage{overpic}
\usepackage{subcaption}
\usepackage{bibtopic}
\usepackage{subeqnarray}
\usepackage{xcolor}
\let\normalcolor\relax
\usepackage{graphicx}
\usepackage{amsmath}
\usepackage{longtable}
\usepackage{color}

\allowdisplaybreaks

\newcommand{\exclude}[1]{}
\def\subsubsubsection{\paragraph}

\def\nn{\nonumber}

\def\d{\mathrm{d}}

\def\a#1{\alpha_{#1}}

\def\beq{\begin{equation}}
\def\eeq{\end{equation}}
\def\be{\begin{equation}}
\def\ee{\end{equation}}
\def\bea{\begin{eqnarray}}
\def\eea{\end{eqnarray}}
\def\bal{\begin{align}}
\def\eal{\end{align}}

\def\2b2[#1,#2][#3,#4]{\left( \begin{array}{cc} #1 & #2 \\ #3 & #4 \end{array}
\right)}
\def\3b3[#1,#2,#3][#4,#5,#6][#7,#8,#9]{\left( \begin{array}{ccc} #1 & #2 #3 \\
#4 & #5 & #6\\#7&#8&#9\end{array} \right)}

\newcommand\fverb{\setbox\pippobox=\hbox\bgroup\verb}
\newcommand\fverbdo{\egroup\medskip\noindent%
                        \fbox{\unhbox\pippobox}\ }
\newcommand\fverbit{\egroup\item[\fbox{\unhbox\pippobox}]}

\newcommand{\bear}{\begin{eqnarray}}

\newcommand{\eear}{\end{eqnarray}}

\newcommand{\bsea}{\begin{subeqnarray}}
\newcommand{\esea}{\end{subeqnarray}}
\newbox\pippobox

\def\f{\varphi}
\def\d{\delta}
\def\g{\gamma}

\def\6{\partial}
\def\a{\alpha}
\def\nn{\nonumber}

\def\pa{\partial}

\def\e{\epsilon}
\def\m{\mu}
\def\n{\nu}
\def\r{\rho}
\def\s{\sigma}

\def\sp{\;\;\;,\;\;\;}

\def\z{\zeta}
\def\sq
\def\a{\alpha}
\def\b{\beta}
\def\l{\lambda}

\def\hri#1#2{\href{http://arxiv.org/abs/#1}{[ArXiv:#1]#2}}
\def\hre#1#2{\href{http://arxiv.org/abs/#1/#2}{[ArXiv:#1/#2]}}
\def\hrj#1#2{\href{https://doi.org/#1}{#2}}

\def\e{\epsilon}

\def\d{\delta}

\def\D{\Delta}

\def\EE{{\cal E}}
\def\FF{{\cal F}}

\def\NN{{\cal N}}
\def\OO{{\cal O}}

\def\ar{~~~\Rightarrow~~~}

\title{Phases and Phase transitions of $U(1)\times SU(2)$ symmetric holographic matter}
\author{
 M. J\"arvinen$^\sharp$, E. Kiritsis$^\natural$$^\flat$$^\dagger$,  F. Nitti$^\natural$, E. Pr\'eau$^\ddagger$
~\\
~\\
$^\sharp$ \href{https://www.apctp.org}{Asia Pacific Center for Theoretical Physics}, Pohang 37673, Republic of Korea, 
and
 \href{https://pheng.postech.ac.kr/}{Department of Physics}, Pohang University of Science and Technology, Pohang 37673, Republic of Korea

\vspace{3mm}
 
$^\natural$ \href{http://www.apc.univ-paris7.fr}{Universit\'e  Paris Cit\'e, CNRS, Astroparticule et Cosmologie,  F-75006 Paris, France}

\vspace{3mm} 

$^\flat$ \href{http://hep.physics.uoc.gr}{Crete Center for Theoretical Physics}, Institute for Theoretical and Computational Physics,
Department of Physics,  P.O. Box 2208,\\
University of Crete, 70013, Heraklion, Greece

\vspace{3mm}
 
$^\dagger$ \href{https://www.theorie.physik.uni-muenchen.de/}{Arnold Sommerfeld Center for Theoretical Physics}, Ludwig-Maximilians-Universit\"at M\"unchen, 80333
M\"unchen, Germany

\vspace{3mm} 

$^\ddagger$\href{https://www.uu.nl/en/research/institute-for-theoretical-physics}{Institute for Theoretical Physics, Utrecht University, 3584 CE Utrecht, The Netherlands}
}
\preprint{APCTP Pre2024 - 019\\
CCTP-2024-15\\
ITCP-2024/15}
\abstract{The phase diagram and symmetry breaking patterns of a holographic CFT with $U(1)\times SU(2)$ symmetry are analyzed using the simplest holographic action, namely Einstein-Yang-Mills (YM) theory with a negative cosmological constant.  This is relevant for both condensed matter and QCD applications. With a U(1) and an ``isospin'' chemical potential turned on, we determine all possible symmetry breaking patterns, which are associated to the condensation of spin-one order parameters. The possible IR asymptotics of the Einstein-YM solutions are derived analytically, both for 2+1 and 3+1 boundary dimensions.
The competing solutions are then computed numerically, both at zero and non-zero temperature, from which the full three-dimensional phase diagram is determined.
We find a surface of second order phase transitions that separate uncondensed and condensed  phases. In some regions with a large fraction of charged to neutral degrees of freedom, the phase transition becomes first order.
}

\begin{document}
\maketitle

\section{Introduction}

\label{Intro}

The gauge/gravity duality provides a way to tackle questions about strongly coupled systems which are otherwise very hard (and sometimes impossible) to attack with other methods. This approach has been widely used in the description of theories relevant for high energy physics (notably QCD, but also possible  strongly-coupled sectors beyond the Standard model). Moreover, several applications to low-energy, strongly-coupled condensed matter systems at quantum critically (famously, high-$T_c$ superconductors and non-Fermi liquids)  have  been proposed \cite{Cubrovic:2009ye}-\cite{Hartnoll:2016apf}

Beyond the realm of high-energy physics, applications to any given specific systems may be complicated. {This is} especially {true} when the details of the microscopic degrees of freedom are important, and {may differ a lot from} field theories {that admit} a gravitational dual, typically large-$N$ gauge theories.

The previous paragraph may give the impression that any practical application of holography to condensed matter physics may be unlikely. However, there are aspects of the duality which are  universal, and independent on the underlying degrees of freedom. One example are the aspects  which reflect the structure of symmetries and their breaking, that control  universal phenomena at strong coupling both in equilibrium (phase structure, phase transitions) and out of equilibrium in a hydrodynamic expansion (which is based on the long-wavelength dynamics of conserved charges).

In this respect, the holographic approach is conceptually similar to a strongly-coupled version of the Landau paradigm. In {the Landau framework},  scaling regimes are controlled by a universal weakly-coupled field theory, which is constructed based on the nature of order parameters that control symmetries and their breaking. Similarly, in holography, symmetries of strongly coupled systems are implemented in a universal way (which we review bellow) in a {\em weakly-coupled} gravitational dual theory, which has easily-identified order parameters related to symmetry breaking. 

The possible thermodynamic phases (at finite temperature and density) are classified  by universal features of the dual geometries, such as whether they have a horizon, whether the solution has non-zero charges, and which order parameters are turned on. The  dynamics governed by the field equations for the bulk fields determines which is the dominant phase at any given value of the temperature, chemical potential and other external control parameters of the model which may encode  microscopic details of the dual field theory. This is  as good as what we obtain  from Landau-Ginzburg theory for weakly coupled systems, which can predict universal behavior close to a phase transition but not e.g. the   value of the critical temperature of a specific realisation.

Moreover, in the holographic framework, it is much easier to describe both transport and far out of equilibrium dynamics, \cite{Zaanen:2015oix,Ammon:2015wua,book}.

\subsection{Symmetries and their breaking in holography}

In holography, global symmetries of the system are realized as gauge symmetries in the gravity dual, with the same gauge group. This means that any exact symmetry will be associated to a propagating spin-one field $A_M^a(z, x^\mu)$ in the bulk, where we denote the $d$ boundary directions with $x^\mu$ and the holographic direction\footnote{For simplicity we ignore in this discussion other compact directions, which  exist in 10-dimensional string theory realisations but do not play any role in this general discussion} by $z$.  If the symmetry on the field theory side is exact, the bulk action must be  gauge-invariant and,  to lowest order in  a derivative expansion, it is uniquely determined to be the Yang-Mills action, with minimal couplings to   charged fields which may exist in the bulk.

The holographic dictionary identifies the parameters which control breaking of the symmetry with expansion coefficients  
near the boundary of the gravitational solution. This relation is simplest when the boundary has the structure of an asymptotic AdS spacetime, where the bulk metric takes the approximate form:
$$
ds^2 \sim {1\over z^2}\left(dz^2 + dx_\mu d x^\mu \right) \qquad z \to 0.
$$
This limit  corresponds to  the dual field theory reaching a ultraviolet (UV)  conformal fixed point,  which we assume here to be the case\footnote{This can be generalized to theories exhibiting  violation of scaling and/or anisotropic behavior \cite{Charmousis:2010zz}.}. In  this case, the gauge-fields behave near the boundary as:
\be \label{in1}
A_\mu^a (x,z) \sim a_\mu^a(x) + v_\mu^a(x) z^{d-1} + \ldots \qquad z\to 0
\ee
where $d$ is the dimension of the spacetime on which the field theory lives. The quantities  $a_\mu(x)$  and  $v_\mu(x)$ are the {\em source} and {\em vev} terms in the dual field theory\footnote{For even boundary dimensions there are log terms in the expansion and a few slight differences, but we do not consider this in the introduction.} :
\begin{itemize}
\item A nontrivial  $a_\mu(x)$  indicates that the  field theory is deformed by adding a term to the action of the form
\be \label{in2}
S_{QFT} \to S_{QFT} + \int d^d x \,  a_\mu^a(x) J^{a\,\mu}(x)
\ee
where $ J^{a\,\mu} $ is the covariantly conserved Noether current generating the symmetry. Thus,  $a_\mu^a(x)$ plays the role of a set of (generalized) chemical potentials for conserved  charge/current densities in the boundary theory;
\item  The term  $v_\mu^a(x)$ plays the role of the vacuum expectation value for the current operator,
\be \label{in3}
 v_\mu^a \propto \langle J^{a}_\mu \rangle
\ee
In particular, a non-zero  $v_\mu^a$ indicates the the solution corresponds to a state with a non-trivial charge/current density turned on.
\end{itemize}
Similarly, any bulk  field $\varphi$ which transforms non-trivially under the gauge group will be dual to a charged operator $O$ in the boundary theory and  have a similar expansion to (\ref{in1}) near the boundary:
\be \label{in4}
\varphi (x,z) \sim j(x) z^{d-\Delta} + v(x) z^{\Delta} + \ldots \qquad z\to 0
\ee
where $\Delta$ is the conformal dimension of $O$, and $j$ and $v$ play the role of the source and the condensate of the operator $O$. In particular, $j$ controls the {\em explicit} breaking of the global symmetry , whereas $v$ plays the role of an order parameter for the {\em spontaneous} breaking if $j=0$. A bulk  solution such that  $v\neq 0$ and $j=0 $ corresponds to  a  broken phase for the symmetry where the charged operator $O$ condenses. This goes along with a spontaneous breaking of bulk gauge-invariance, and a ($z$-dependent) mass for the gauge field $A_M$.

The equilibrium phase diagram of the theory in the grand-canonical ensemble (i.e. as a function of temperature and the various charge/current chemical potentials $a^a_\mu$) is determined by finding all the possible bulk solutions which are compatible with the  boundary conditions determined by  $a^a_\mu$ at leading order as in (\ref{in1}), and which satisfy appropriate regularity conditions in the interior. These conditions fix the condensates, typically to a discrete set of values. The   solution which dominates the ensemble is the one with lowest free energy, which is computed by the bulk on-shell action.

In the case of a non-abelian symmetry, the components of the non-abelian currents may play the role of order parameters, since they transform non-trivially. In this case the symmetry is broken by a spin-one condensate. On the gravity side, this corresponds to the condensation of a (spatial) component of the bulk non-abelian gauge fields. This is the case of interest in this paper.

\subsection{A simple holographic realization of  global  $U(1)\times SU(2)$ symmetry}

In this work, we shall study the general patterns of  symmetry breaking, and the corresponding phase diagram,  in a holographic model which corresponds to a conformal field theory with a global  $U(1) \times SU(2)$ symmetry. We shall consider the minimal dynamics described by the metric and the gauge fields dual to the symmetry currents. In the absence of any other ingredients (charged operators, corresponding to other bulk fields), the symmetry currents themselves will play the role of condensing fields (order parameters).  We study the phase diagram resulting from turning-on the most general combination of simultaneous chemical potentials in  the Cartan subalgebra\footnote{In particular, our results will apply to the subcases when one has only $U(1)$ or only $SU(2)$ symmetry groups.} i.e.  for the $U(1)$ charge and for the ``isospin'' charge corresponding to the  $\sigma^3$ generator  in  $SU(2)$.

The   symmetry structure we consider, can be found for example in strongly coupled neutral or U(1)-charged  non-relativistic spin systems (e.g. cold atoms) as well as strongly coupled materials  with vector-like order parameters (p-wave superconductors). One expects that the universal feature of symmetry breaking in these models may be captured by holography (see e.g. \cite{Zaanen:2015oix,Ammon:2015wua} for gauge/gravity duality applications to condensed matter physics).
Moreover, modulo a slight generalization that we discuss below, this symmetry is relevant for {\em nuclear and particle physics}, in particular for the dense state of nuclear matter found in neutron stars.

Indeed, the  $U(1) \times SU(2)$ group is the  vector part of the flavor group of  QCD with 2 quarks, which is $U(2)_L\times U(2)_R$. The non-abelian axial  subgroup is broken spontaneously by chiral condensation  and explicitly by quark masses, and the axial $U(1)_A$ is broken  by the axial anomaly. The surviving vector subgroup  $U(1)_B\times SU(2)_I$ corresponds to baryon number and isospin which, in the approximation that $m_u\simeq m_d$ and if we disregard the electroweak interactions, is an exact symmetry of QCD with two flavors.

The dense state found in neutron stars, is characterized by a non-zero baryon charge  and an isospin imbalance (neutrons are more abundant than protons). This corresponds  precisely to the two (electric) chemical potentials in the two commuting  subgroups $U(1)_B$ and $U(1)_I \subset SU(2)_I$ of the global symmetry group.

The simplest  holographic model which realizes this symmetry is a  $(d+1)$-dimensional bulk theory where Einstein gravity with a negative cosmological constant is coupled to Yang-Mills theory with gauge group $U(2)_L\times U(2)_R$. The action we use reads, schematically\footnote{{The precise definition of the holographic model is given in section} \ref{Sec:HM}.},
\begin{equation} \label{in5}
S =  M_p^{d-1}\int \sqrt{g} \Big[ R - 2\Lambda + w_0^2 \Big( \mathrm{Tr} F^2_L + \mathrm{Tr} F^2_R \Big) \Big] \, ,
\end{equation}
where $w_0$ is the inverse gauge coupling and $F_{L,R}^{MN}$ are Yang-Mills field strengths for the left and right gauge fields $A^M_{L,R}$. The action (\ref{in5})  contains all the lowest-dimension operators consistent with the symmetries and the field content we study, and is valid in a low-energy approximation. It is expected to be corrected both by higher curvature terms and by terms of higher order in the field strengths, which typically take the form of a DBI action.  Another  ingredient which is  present in string-derived and string-inspired holographic flavor actions (in odd bulk dimensions), and that we are not including in (\ref{in5}), are  Chern-Simons terms. These  give no contribution to the homogeneous phases we study in this paper, but they may affect possible non-homogeneous phases as well as transport. An example of such effects in the context of dense QCD has been recently discussed in \cite{CruzRojas:2024igr,Demircik:2024aig}.

For our study of symmetry breaking and the phase diagram,  we concentrate on the vector part of the gauge fields and suppose that the axial part is trivial, setting $A_L^M= A_R^M$ . This reduces (\ref{in5}) to a single $U(1)_B\times SU(2)_I$ Einstein-Maxwell-Yang-Mills action.  The simplest possible phase with non-zero baryon and isospin chemical potential,  is one where only baryon  and isospin  charges condense, and all gauge fields except the time-components of the Maxwell field  $A_0^B$ and of the third Isospin gauge field  $A_0^3$ are trivial. The corresponding solution is an AdS-Reissner-Nordstr\"om (AdS-RN) black hole,  carrying both baryon and isospin charge.

As was already discussed by Gubser and collaborators in the seminal works \cite{Gubser08a,Gubser08b} for a pure $SU(2)$ theory,  other solutions exist which features the condensation of the spatial  components of the gauge field components along the $\sigma^1$ and $\sigma^2$  generators of $SU(2)$. These are ``superconducting'' solutions, in the sense that they have a current turned on, which is not supported by a corresponding electric field.  In certain regions of parameter space (in particular, for small $w_0$ and at low temperature) these  dominate over the AdS-RN solution.

Taken at face value, as a model for flavor physics,  (\ref{in5}) is   oversimplified: it only captures the unbroken symmetries of the QCD vacuum in the flavor sector but not much more. For one thing, the theory we use is conformally invariant. It misses all the dynamics of the glue sector and of the running gauge coupling,  which breaks conformal symmetry and gives rise to the dynamical non-perturbative mass scale in Yang-Mills and QCD. Moreover, the model does not include the dynamics leading to chiral symmetry breaking. In holographic models the description of these features requires other fields and a more complicated construction: at least a scalar field dual to the color field strength\footnote{{More precisely to the gauge-invariant operator $\mathrm{Tr}(G^2)$, with $G$ the color field strength.}}, and an additional matrix of  scalar fields transforming in the bi-fundamental representation of the flavor group (which here is  $U(2)_L\times U(2)_R$ but can be taken more generally to be  $U(N_f)_L\times U(N_f)_R$ to account for $N_f$ quark flavors). These features are  included in more complex models, such as the top-down Witten-Sakai-Sugimoto model from type IIA string theory \cite{SS}, or bottom-up models such as V-QCD \cite{V1,V2,V3}. However, we expect that the qualitative features of the solutions obtained in the simplified model (\ref{in5}), if not the details,  will carry over to these more complete models: in particular, the classification of the order parameters and the phases of the theory does not depend on the presence of other fields or on the detailed form on the metric. What will change will be the location of the phase transitions between different solutions.

Even with these simplifications, the model presented here can still be interesting in connection with dense QCD matter, in at least two ways:
\begin{enumerate}
\item  It can provide a qualitative scan of  the  set of possible   (deconfined) phases  which may be potentially relevant for the core of neutron stars\footnote{For related studies of isospin asymmetry in holographic models in the context of dense matter and neutron stars, see \cite{Kim:2010dp}-\cite{Bartolini:2023oxs}.},  including those with condensation of vector-like order parameters, which may lead to different equations of state for the core, {and, especially, different transport properties};
\item It would provide a more general stage than what has been considered so far for the calculation of transport of weak currents in  holography along the lines of \cite{neutrino}: in that work, weak  correlators relevant for neutrino transport were computed holographically in the simplest AdS-RN toy model with non-zero baryon chemical potential but no isospin imbalance. It is possible that, once we introduce an isospin chemical potential, the AdS-RN black hole in  which that computation was performed may not be the dominant phase: depending on the value of $w_0$, it is possible that the true ground state of the boundary theory  is in a superconducting phase\footnote{In QCD it is known that turning on an isospin chemical potential, eventually forces the $\pi^{\pm}$ to condense,  \cite{Son}. At higher values, $\rho^{\pm}$ gauge bosons are expected to condense, as was seen in holographic models, \cite{Parnachev}-\cite{Kovensky}.}.   Even before carrying the computation to more realistic theories it important to understand how adding an isospin chemical potential affects both the background and the correlators.
\end{enumerate}

\subsection{First steps in the phase diagram: Gubser's pioneering work}

The idea that, in  holographic $SU(2)$ Einstein-Yang-Mills theory, a chemical potential in the $\sigma^3$ direction can trigger spontaneous condensation of the $\sigma^1$ and $\sigma^2$ gauge fields is due to Gubser, which in \cite{Gubser08a} extended the concept of holographic superconductor \cite{Gubser:2008px,Hartnoll:2008vx} to the case of vector order parameters. In that work and its follow up   \cite{Gubser08b} with Pufu, Gubser considered a holographic model  of $(2+1)$-dimensional field theories with a global $SU(2)$ symmetry, dual to $(3+1)$-dimensional Einstein-Yang-Mills with gauge group $SU(2)$.  Both works have shown the onset of a (isospin-)superconducting background, competing with the AdS-RN solution,  where spatial components of the charged current have a non-vanishing expectation value and break isospin symmetry completely. These solutions were found to exist for sufficiently large values of the bulk Yang-Mills coupling (controlled by $w_0^{-1}$ in (\ref{in5})), and to dominate over the RN solution at large isospin chemical potential (in units of the temperature).

More specifically, in   \cite{Gubser08a} Gubser considered a class of solutions with non-zero isospin chemical potential in the $\sigma^3$-direction,  and  non-trivial spatial components of the bulk gauge field  along one of the other isospin components, in such a way that the solution breaks the isospin and spatial rotations while preserving a combination of the two, corresponding to an $SO(2)$ subgroup. The solutions and the phase diagram  were constructed numerically.

In  \cite{Gubser08b}, a different class of solutions  was considered, this time breaking  completely the (isospin)$\times$(rotations) symmetry.  In this case the analysis was carried out in the  limit of large gauge coupling constant, in which one can treat the gauge fields as perturbations over a fixed AdS-RN metric. By analyzing the effective mass of the gauge fields in the near-horizon $AdS_2$ region of the AdS-RN metric, Gubser and Pufu argued for  the presence of an instability driving the theory to the isospin-superconducting phase.

These works did not study the full phase diagram, containing both classes of solutions and possibly others with different symmetry breaking patterns (as well as the AdS-RN black hole).  In particular the question whether the condensed phase starts dominating exactly at the onset of $AdS_2$ instability, or at a higher temperature (and/or smaller value of the Yang-Mills coupling) was left open.

\subsection{A walk through the full  $U(1)\times SU(2)$ phase diagram: summary of results}

In this work, we investigate the phases of the simple holographic model with the action \eqref{in5}, generalizing the analysis of Gubser and Pufu reviewed above. We work in the canonical ensemble, and consider the phase space parametrized by three dimensionless parameters: the ratio of temperature to isospin density $T/n_3^{1/3}$, the ratio of isospin to quark number density $n_3/n_q$ and the (inverse) flavor coupling $w_0$.

The ways we go beyond the previous works are the following
\begin{itemize}

\item  We add an extra U(1) to the SU(2) of Gubser, and we turn on an extra associated chemical potential.
As we shall see in some cases this changes the story importantly.

\item We do a symmetry analysis to determine all vev possibilities up to symmetry.

\item By an asymptotic analysis of the Einstein-YM equations in the IR,  we classify all IR fixed points, and subsequently all possible solutions to the equations, using the techniques of \cite{Charmousis:2010zz} and especially \cite{Gouteraux12}.

\item We find all competing solutions always including the full back-reaction.

\item We find all competing solutions at zero temperature.

\item  We do the analysis both in 2+1 and in 3+1 boundary dimensions.  

\end{itemize}

In section \ref{OP}, we start by determining what are the order parameters that distinguish the various phases. At high temperature, the thermal state is described by the AdS-RN background with both baryon number and isospin charge\footnote{The explicit form of the RN solution is reviewed in section \ref{Sec:bsT} .}. This state is invariant under the group SO($d-1$)$\times$U(1)$\times$U($N_f - 2$), where the first factor is the group of spatial rotations, and the remaining two correspond to the isospin subgroup preserved by a finite isospin density. At low temperature, a (flavor) superconducting condensate may form, that breaks spontaneously part of this symmetry group.

There exist several superconducting phases, depending on which component of the isospin current condenses. As explained in section \ref{OP}, the singular value theorem makes it possible to classify all the condensation channels, for any dimension $d$ and number of flavors $N_f$. They all correspond to p-wave superconductors, and differ by the symmetries that they leave unbroken. In particular, for the case that we consider of $N_f = 2$ flavors and $d \geq 2+1$ boundary dimensions, there exist two types of condensates that we label (0,1) and (1,1)\footnote{The origin of this labelling comes from the structure of solutions in the bulk, and is explained in section \ref{OP}.}
\begin{itemize}
\item The (0,1) phase is such that a single component of the isospin current (say $J^1_z$) condenses. This phase preserves an SO($d-2$) subgroup of SO($d-1$) and breaks U(1);
\item In the (1,1) case, two components condense together with the same amplitude (say $J^1_x$ and $J^2_y$), preserving an SO(2)$\times$SO($d-3$) subgroup of SO($d-1$)$\times$U(1). 
\end{itemize}

The ans\"atze associated with the (0,1) and (1,1) phases are presented in the main text, in equations \eqref{PD25}-\eqref{PD28}. Based on these, the phase diagram of the model can be computed numerically, along the lines of the general method reviewed at the beginning of section \ref{Sec:PD}. We first summarize the results for the full parameter space at finite temperature, and then discuss in more details the limit of zero temperature.

\subsubsection*{The phase diagram at finite temperature}
\addcontentsline{toc}{subsubsection}{The phase diagram at finite temperature}

We consider two cases for the number of boundary dimensions, starting with $d=2+1$ in section \ref{Sec:PD2d}. In this case, the set-up is precisely the same that was considered by Gubser in \cite{Gubser08a}, so our analysis is a direct extension of his work. Specifically, we provide a complete description of the phase diagram in the plane of isospin density and Yang-Mills coupling, including the (0,1) phase that \cite{Gubser08a} did not take into account. Our results indicate that (0,1) actually always dominates over (1,1) (see figure \ref{fig:Dffull3d}), which has interesting consequences for the phase diagram (figure \ref{fig:PTl3d}). That is, whereas including only (1,1) predicts a second order phase transition, the transition to the (0,1) phase becomes first order at large $w_0$ (small Yang-Mills coupling).

Some of these features were observed in previous works: in \cite{Ammon09}, the (0,1) solution was shown to dominate over (1,1), whereas \cite{Arias12} found that the transition to (0,1) becomes first order at large $w_0$. Our work improves on those results by first justifying that there are no other competing solutions beyond (0,1) and (1,1), and second providing the full phase diagram over the entire range where the paired phase exists.

Section \ref{Sec:PD3d} discusses $d=3+1$ boundary dimensions, which is the main focus of our work.
In this case, we allow for a finite quark number density $n_q$, such that a typical point in the parameter space corresponds to some kind of quark matter with isospin imbalance. The resulting three dimensional phase diagram is shown in figure \ref{fig:PDnq4d}. As for the three-dimensional case, the leading superconducting phase is always (0,1), and the transition becomes first order at large $w_0$ and $n_3/n_q$. When $n_3/n_q$ is decreased, the condensed phase covers a smaller and smaller portion of the phase diagram, until it completely disappears at $n_3 = 0$ (as it should). Interestingly, the equations for the quark number gauge field are simple enough, that an exact expression \eqref{d41} can be derived for the $n_q$-dependence of the phase transition surface, in the regions where it is second order.

The phase diagram of figure \ref{fig:PDnq4d} allows to answer one of the questions formulated in the introduction, that is whether an isospin superconducting phase may be relevant to neutron-star-like conditions. Imposing the physical constraints of charge neutrality and $\beta$-equilibrium (see appendix \ref{CN}), we find a stringent constraint on the ratio of isospin to quark number density
\begin{equation}
\label{in6} \frac{|n_3|}{n_q} \leq \frac{1}{6} \, .
\end{equation}
In this regime - delimited by the purple line in figure \ref{fig:PDnq4d} - only a very small part of the parameter space at low temperature and $w_0$ lie in the condensed phase. The corresponding values of $w_0$ are an order of magnitude smaller than the numbers (of order 1) that arise naturally from top-down extensions of the model, or from fitting to lattice QCD data. Hence, our result for the phase diagram indicates that the p-wave condensation is unlikely to occur in physical conditions relevant for quark matter, even in the limit of very small temperature like in a neutron star.

\subsubsection*{Zero-temperature solutions}
\addcontentsline{toc}{subsubsection}{Zero-temperature solutions}

The bulk solutions that arise in the limit of vanishing temperature are qualitatively different from their finite temperature counterparts. In the last section \ref{Sec:T0}, we provide a detailed analysis of the zero-temperature solutions. This analysis completes our construction of the phase diagram, and gives interesting insight into the physics of cold quark matter in the model \eqref{in5}.  

In the uncondensed RN phase, the zero temperature limit is trivially given by the extremal black hole. The non-trivial problem is to compute (numerically) the superconducting (0,1) solution at $T=0$. To do this, it is necessary to first determine the infrared (IR) regularity conditions associated with the solution, which differ from those at finite temperature. The appropriate ansatz for the IR conditions was found by observing the qualitative behavior of the (0,1) solution in the near-extremal limit\footnote{In appendix \ref{G}, it is checked that there are no other IR asymptotics corresponding to regular solutions.}. The result is such that the bulk fields $\varphi$ follow a double series near the extremal horizon
\begin{equation}
\label{in7} \f(u) = u^{c_\f}\sum_{j,k \in \mathbb{N}} \f_{jk}(c_3) u^{j\, \a(c_3)} u^{k\, \g(c_3)} \, ,
\end{equation}
where $u$ goes to 0 at the horizon, and the powers are  generically non-integer. The coefficients and the exponents of the expansion depend on a single parameter $c_3$, which controls the density ratio $n_3/n_q$.

Interestingly, the form of the series expansion \eqref{in7} was found to depend on the value of $c_3$, with two branches separated by the point $c_3 = 3/2$. In particular, the exponents are given by
\begin{equation}
	\label{in8} \a(c_3) =
	\left\{
	\begin{array}{rl}
		2c_3 - 2 & \, ,\,\, 1< c_3 < \frac{3}{2} \\
		1 & \, ,\,\, c_3 > \frac{3}{2}
	\end{array}
	\right. \sp \g(c_3) = \left|2c_3 - 3\right| \, .
\end{equation}
The first few terms of the expansion in each branch are shown in \eqref{T0n5}-\eqref{T0n10}. Even though the numerical results shown in appendix \ref{Thc32} do not show any sign of a finite order phase transition across $c_3 = 3/2$, we did observe a qualitative difference in the low temperature behavior of the specific heat
\begin{equation}
\label{in9} C_V \equiv T \frac{\partial s}{\partial T} \propto T^{a(c_3)} \sp
\left\{
\begin{array}{rl}
	a(c_3)<1 & \, ,\,\, 1< c_3 < \frac{3}{2} \\
	a(c_3) = 1 & \, ,\,\, c_3 > \frac{3}{2}
\end{array}
\right. .
\end{equation}
A plausible scenario is that an infinite order BKT-like transition may happen at $c_3=3/2$.

With the IR expansions \eqref{T0n5}-\eqref{T0n10}, we were able to compute numerically the solution corresponding to the (0,1) superconductor at zero temperature. The solution has the following properties:
\begin{itemize}
\item It contains an extremal horizon, which is charged under baryon number but not under isospin. This horizon has a non-abelian hair, which, unlike the horizon, is charged under isospin but not baryon number. From the field theory point of view, this means that the baryon number is fractionalized (contained in unconfined quarks), whereas isospin is carried by color singlet bound states - the quark pairs that constitute the p-wave condensate;
\item When the second order (quantum) phase transition is approached, the bulk fields tend to the RN solution. This approach is mostly smooth, except for the non-abelian hair which is peaked at the horizon. This behavior is associated with the bulk isospin density tending to a Dirac delta at the transition, which is located on the extremal horizon;
\item The geometry of the solution near the extremal horizon is AdS$_2\times \mathbb{R}^3$, which means that the dual theory is effectively described by a CFT$_1$ at low energies. This is the same geometry that arises in the IR of the RN phase, but with different effective sources at the AdS$_2$ boundary. The RN instability that triggers the phase transition to (0,1) is thus non-normalizable in the IR AdS$_2$. This is why the quantum transition is second order, and not infinite order, as happens for an AdS$_2$ instability \cite{Jensen10,Iqbal10,V1b,V6};
\item The IR operator dual to the isospin density becomes irrelevant in the paired phase, and its conformal dimension increases as the condensate grows. This is a manifestation of the screening of the isospin charge in the condensed phase.
\end{itemize}

The zero-temperature (0,1) solution changes qualitatively in the limit where $n_3/n_q$ goes to infinity. In this case, the appropriate regular behavior for the bulk fields $\f$ in the IR is given by an exponential scaling
\begin{equation}
\label{in10} \f(r) \underset{r\to\infty}{=} \sum_{j \in \mathbb{N}} \f_j(r) \ex^{-j\, d_\f r} \, ,
\end{equation}
where $r\to\infty$ is the IR limit. The first few terms in this expansion are given by \eqref{T05}-\eqref{T09}.

In this regime, the solution is horizon-less, and the IR geometry is AdS$_5$. The solution therefore represents a renormalization group flow from a UV fixed point to an IR fixed point, both described by a four-dimensional CFT.

The behavior of the solution close to the phase transition is also different from the case of finite $n_3/n_q$, since the transition is now first order. Specifically, at the point where the two phases become degenerate, the geometry is arbitrarily close to RN, but the non-abelian hair still extends over a significant distance away from the (emergent) horizon. When the first order transition happens, it is realized by the gravitational collapse of the hair.

\subsection{Open questions and future directions}

There are some immediate questions that are of interest:

\begin{itemize}

\item The generalization to higher $N_f$ is not completely straightforward, with the case $N_f=3$ of potential phenomenological relevance. We suspect that it is always the equivalent of the (0,1) phase that dominates (single component condensates), but it would be very interesting to verify this.
Moreover, the $N_f>2$ case allows in principle a richer set of isospin-like chemical potentials to be turned on, and this can produce a more complicated phase diagram. 
    
\item An action for the gauge fields that is different from the YM action may be envisaged. Generically we may expect a DBI-like action. It is not entirely clear whether this may allow new phases, or that it has only quantitative impact on the phase diagrams we found.

\item The presence of a general $U(1)\times SU(2)$ Chern-Simons term adds one more parameter to the theory, but does not modify the translationally-invariant solutions. It allows however for solutions that break spontaneously the translational symmetry (density waves or chequered phases) \cite{OP09,OP10}, which may compete with the translationally-invariant ones. This analysis is very interesting and may have phenomenological implications.

\item The general method for determining the most general order parameters may be applied to other set-ups with isospin in holographic QCD, with richer phase structure, including baryonic matter and chiral symmetry breaking.
    
\item The calculation of current-current correlators in the condensed phases is an interesting problem. It is more complicated than in the unbroken phases~\cite{neutrino,Hoyos:2024pkl}, but may be phenomenologically relevant to transport in dense matter.

\end{itemize}

\section{The holographic model}

\label{Sec:HM}

We introduce in this section the holographic model that will be  used to describe flavor physics.
It is the simplest bottom-up model describing the dynamics of chiral current operators.

We assume that the medium is described by a strongly interacting quantum  theory with $N_f$ flavors and $U(N_f)_L\times U(N_f)_R$ chiral symmetry.
If the field theory lives in four dimensions, according to the holographic duality, this theory is dual to a five-dimensional gravitational theory.
We simplify the dynamics of the glue, assuming it is conformal. Then the gravitational ground state solution  will be given by five-dimensional Anti-de Sitter space AdS$_5$, which is a constant negative curvature space with a  four-dimensional boundary.

\subsection{Action}

We consider  an 
asymptotically AdS bulk theory, whose field content is dictated by the types of operators that we want the dual (boundary) quantum field theory to include. In the present case, the operators of interest are the chiral currents $J^{(L/R)}_\m$, which are dual to chiral gauge fields in the five-dimensional bulk $\mathbf{L}_M$ and $\mathbf{R}_M$. The latter, are elements of the Lie algebra of the chiral group U($N_f$)$_L\times$ U($N_f$)$_R$. The bulk gravitational action is constructed as the sum of a color and a flavor part
\be
\label{Sb} S = S_c + S_f \, .
\ee
The action for the color sector is the $(d+1)$-dimensional Einstein-Hilbert action
\be
\label{SEH} S_{\text{c}} = M^{d-1}N_c^2 \int\mathrm{d}^{d+1}x\sqrt{-g}\, \left(R + \frac{d(d-1)}{\ell^2}\right) \, ,
\ee
where $R$ is the Ricci scalar, $M$ the Planck mass, $\ell$ the AdS radius and $N_c$ the number of colors.  For the flavor sector, we make the simplest choice of a quadratic Yang-Mills action for the chiral gauge fields
\be
\label{SYM} S_f = - \frac{1}{8\ell^{d-3}}(M\ell)^{d-1}w_0^2 N_c \int\mathrm{d}^{d+1}x\sqrt{-g} \left(\text{Tr}\,\mathbf{F}_{MN}^{(L)}\mathbf{F}^{MN,(L)} + \text{Tr}\,\mathbf{F}_{MN}^{(R)}\mathbf{F}^{MN,(R)} \right) \, ,
\ee
where $w_0^{-1}$ is proportional to the flavor Yang-Mills coupling, and $\mathbf{F}^{(L/R)}$ are the field strengths of the gauge fields $\mathbf{L}$ and $\mathbf{R}$
\be
\label{defF} \mathbf{F}^{(L)} \equiv \mathrm{d}\mathbf{L} - i\mathbf{L}\wedge \mathbf{L} \sp \mathbf{F}^{(R)} \equiv \mathrm{d}\mathbf{R} - i\mathbf{R}\wedge \mathbf{R} \, .
\ee

As usual in holographic theories, the number of colors $N_c$ is assumed to be large in order for the semi-classical treatment of the bulk theory to be valid. Since we are interested in describing dense baryonic matter, the back-reaction of the flavor sector on the glue sector will play an important role. In order for this back-reaction to be finite, we consider the so-called Veneziano large $N$ limit
\be
\label{V1} N_c \to \infty \sp N_f \to \infty \sp \frac{N_f}{N_c} \text{ fixed} \, .
\ee
Although $N_c$ and $N_f$ are assumed to be large, finite values of $N_c$ and $N_f$ will eventually be substituted in the large $N$ result for phenomenological applications. Specifically, $N_c$ will be set to 3, and from now on we fix the flavor sector to be composed of $N_f = 2$ massless flavors.  When the chiral group is U(2)$_L\times$U(2)$_R$, the chiral currents and their dual gauge fields can be decomposed in the Pauli basis $\{\s_a\}$
\be
\nn J^{(L)}_\mu = \frac{1}{2}\hat{J}^{(L)}_\mu\mathbb{I}_2 + \frac{1}{2}\sum_{a=1}^3 J^{a,(L)}_\mu \s_a \sp \mathbf{L}_M = \frac{1}{2}\hat{L}_M\mathbb{I}_2 + \frac{1}{2}\sum_{a=1}^3 L^{a}_M \s_a \, ,
\ee
\be
\label{JsP} J^{(R)}_\mu = \frac{1}{2}\hat{J}^{(R)}_\mu\mathbb{I}_2 + \frac{1}{2}\sum_{a=1}^3 J^{a,(R)}_\mu \s_a \sp \mathbf{R}_M = \frac{1}{2}\hat{R}_M\mathbb{I}_2 + \frac{1}{2}\sum_{a=1}^3 R^{a}_M \s_a \, ,
\ee

It is instructive at this point to compare this simplified holographic model, with V-QCD, \cite{V1}, that is expected to contain all the ingredients relevant for the Veneziano limit.
The flavor neutral part, apart from the graviton contains a scalar (with a potential),  that is dual to the $\mathrm{Tr}[F^2]$ operator of YM theory. Its non-triviality in the ground state solution reflects the YM RG flow that breaks the conformal invariance of the UV Theory.
In the  flavor sector, apart from the gauge fields dual to currents, there is the bifundamental tachyon scalar that is dual to the quark mass operator, \cite{ckp}.
The flavor action is given by the non-linear DBI tachyon action \cite{V1}.\

Here we removed both scalars,  we linearized the DBI action and we dropped the tachyon Chern-Simons terms, \cite{V7}.

\subsection{Phases}

In this work, we investigate the (time-independent) solutions of the classical equations of motion derived from the action \eqref{Sb}-\eqref{SYM}. Those solutions, if stable, describe possible equilibrium states in the boundary theory. We are  interested in states at finite temperature $T$, quark number chemical potential $\m_q$ and isospin chemical potential $\m_3$. The two chemical potentials correspond to non-trivial sources of the gauge fields, while the temperature controls the vev of the energy momentum tensor.

Exhibiting the possible solutions as a function of the dimensionless parameters $(w_0,\m_q/T,\m_3/T)$, and comparing their free energies, makes it possible to construct the phase diagram of the theory in this parameter space. Note that here the parameter $w_0$ plays a slightly different role than the other parameters, as changing its value amounts to modifying the boundary theory rather than its thermodynamic state.

To compute the phases of the model, we need to solve the equations of motion, whose general form is given by the Einstein-Yang-Mills equations
\begin{align}
\nn R_{MN} - \frac{1}{2}\left(R + \frac{d(d-1)}{\ell^2}\right)g_{MN} = -\frac{w_0^2\ell^2}{4N_c} \text{Tr}\bigg\{\mathbf{F}^{(L)}_{\hphantom{(L)}MP} &\mathbf{F}^{(L)P}_{\hphantom{(L)P}N} + \frac{1}{4}\mathbf{F}^{(L)}_{PQ}\mathbf{F}^{(L)PQ}g_{MN}  +\\
\label{PD1} & + (L \leftrightarrow R) \bigg\} \, ,
\end{align}
\be
\label{PD2} D^{(L/R)}_M\Big(\sqrt{-g}\mathbf{F}^{(L/R)MN}\Big) = 0 \, ,
\ee
with $D^{(L/R)}_M$ the Yang-Mills covariant derivatives
\be
\label{PD3} D^{(L)}_M \equiv \partial_M - i [L_M,\, . \,\,] \sp D^{(R)}_M \equiv \partial_M - i [R_M,\, . \,\,] \, .
\ee
The YM equations in components become
\be
\pa_{M}(\sqrt{-g}\hat{F}^{MN}) = 0 \, ,
\ee
\be
{1\over \sqrt{-g}}\pa_{M}(\sqrt{-g}F^{a, MN})+\e^{abc}A^a_MF^{b,MN} = 0 \, ,
\ee
where $A$ stands for either of the gauge fields $L,R$.

Equations \eqref{PD1} and \eqref{PD2} will be solved for a specific ansatz, appropriate for the states that we wish to describe. First of all, we are interested in static and homogeneous configurations, where all the fields depend only on the holographic coordinate $r$, defined such that the boundary is reached at $r=0$. Also, it is convenient to work in the radial gauge, where
\be
\label{PD3b} \mathbf{L}_r = \mathbf{R}_r = 0 \, .
\ee

The chemical potentials $\m_q$ and $\m_3$, respectively source the abelian and non-abelian part of the time-component of the vector gauge field, corresponding to the ansatz
\be
\label{PD4} \hat{L}_t = \hat{R}_t \equiv  \Phi(r) \sp L_t^3 = R_t^3 \equiv  \Phi_3(r) \, .
\ee

The rest of the ansatz will be different for each of the possible phases. In the next subsection, we present the simplest background solution, corresponding to the uncondensed phase. The order parameters for condensation and associated ans\"atze are discussed in section \ref{OP}.

\subsection{The simplest background solution at finite density}

\label{Sec:bsT}

We now present the simplest background solution for the bulk action \eqref{SEH}, at finite temperature and density. The dual state of matter that it describes in the dual boundary theory, corresponds to a plasma of deconfined (generalized) quarks and gluons. Introducing a finite density of deconfined baryonic matter is equivalent to sourcing the bulk baryon number gauge field with a chemical potential
\be
\label{smq} \Phi\big|_{\text{boundary}} = \m_q \, .
\ee

In~\eqref{smq}, $\m_q$ is the quark number chemical potential, related to the baryon number chemical potential by $\mu_B = N_c\mu_q$. Likewise, the isospin asymmetry is introduced by sourcing the isospin gauge field with an isospin chemical potential
\be
\label{sm3} \Phi_3\big|_{\text{boundary}} = \m_3 \, .
\ee
Then, the background solution is given by the solution of the Einstein-Yang-Mills equations \eqref{PD1}-\eqref{PD2} obeying the boundary conditions \eqref{smq} and \eqref{sm3}, together with appropriate regularity conditions in the IR. {The derivation of the solution is reviewed in appendix} \ref{Sec:bgd}. It corresponds to an asymptotically AdS$_{d+1}$ Reissner-Nordström (RN) black-hole\footnote{The phase diagram and the associated thermodynamics of RN black holes has been studied in holography in \cite{RN1,RN2}.}. Specializing to $d=3+1$ dimensions, the metric of this solution reads

\be
\label{ds2RN} \mathrm{ds}^2 = \ex^{2A(r)}\left( -f(r)\mathrm{dt}^2 + f(r)^{-1}\mathrm{dr}^2 + \vec{\mathrm{d x}}^2 \right) \, ,
\ee
where
\be
\label{afRNBH} \ex^{A(r)} = \frac{\ell}{r} \sp f(r) = 1 - \left(\frac{r}{r_H}\right)^4\left(1 + 2\left(1 - \pi T r_H\right)\right) + 2\left(1 - \pi T r_H\right)\left(\frac{r}{r_H}\right)^6  \, ,
\ee
\be
\label{rHRN} r_H = \frac{2}{\pi T} \left( 1 + \sqrt{1 + \frac{w_0^2}{3N_c}\frac{\mu^2}{\pi^2T^2}} \right)^{-1} \, .
\ee
The background solution for the gauge fields is given by
\be
\label{AhA3s2} \Phi(r) = \mu_q\left(1 - \left(\frac{r}{r_H}\right)^2\right) \sp \Phi_3(r) = \mu_3\left(1 - \left(\frac{r}{r_H}\right)^2\right) \, .
\ee
In \eqref{ds2RN}, the coordinate $r$ is the holographic coordinate, defined such that the AdS  boundary is located at $r=0$ and the horizon at $r = r_H$. The total chemical potential that constitutes the charge of the black hole is given by
\be
\label{defmu} \mu \equiv \sqrt{\mu_q^2+ \mu_3^2} \, .
\ee

If we consider conditions relevant for neutron stars, then  the baryon {chemical potential}  is much higher than the temperature, i.e.  $\mu \gg T$. In this limit, the charged black-hole is nearly extremal and the horizon radius is essentially controlled by the chemical potential
\be
\label{rH2} r_H = \sqrt{\frac{3N_c}{w_0^2}}\frac{2}{\mu}\left( 1 + \OO\left(\frac{T}{\mu}\right)\right) \, .
\ee

This RN solution describes a single phase of the associated flavor system at finite density and temperature.
However, as shown already in \cite{Gubser:2008yx,Gubser:2008sz} for 2+1 boundary dimensions, there are other phases where some of the dual operators obtain expectation values that spontaneously break some of the symmetries. We shall discuss such order parameters in the next section.

\section{The order parameters}

\label{OP}

In this section we shall discuss the possible order parameters of the problem and the associated symmetry breaking patterns~\cite{Gubser:2008yx,Gubser:2008sz}. The details of the discussion depend on the dimension. However, the main points are essentially dimension independent. We  first discuss the simplest case with   2+1 boundary dimensions, and then argue how this discussion generalizes to 3+1 and higher number of boundary dimensions.

\subsection{$d=2+1$}

The expansion of the $\Phi$'s \eqref{PD4} near the boundary $(r\to 0)$ is given by
\be
\label{PD5} \Phi(r) = \m_q + \hat{n}_q r + \OO(r^2) \sp \Phi_3(r) = \m_3 + \hat{n}_3 r + \OO(r^2) \, ,
\ee
with $\hat{n}_q$ and $\hat{n}_3$ respectively proportional to the quark and isospin densities.

As both the sources belong to the vector subgroup of the chiral symmetry we should expect that the axial part remains trivial.
Therefore we choose
\be
\mathbf{L}_{\mu} = \mathbf{R}_{\mu}= \mathbf{A}_{\mu}
\label{PD5a}\ee
where $\mu=t, x, y$ are the boundary coordinates.
There is no chiral symmetry breaking in the simplified model we consider, and the condition (\ref{PD5a}) is the closest analogue to this dynamics in the present model. 

The main remaining question is which of the gauge field components $\mathbf{A}^a_{\mu}$, beyond $\Phi = \hat A_t$ and $\Phi_3 = A^3_t$ that are sourced, can have a non-trivial vev.
To answer this question we shall use physical intuition, the bulk Gauss law\footnote{Which is equivalent in the Fefferman-Graham (FG) coordinate system to boundary (covariant) current conservation.}, and symmetry arguments.

None of the bulk gauge degrees of freedom has minimal charge under the $U(1)$ baryon gauge field $\hat A_t$.
Moreover, the $A^{1,2}_{\mu}$ components are minimally charged under the  $A^3_t$ gauge field that is turned on. As this background resembles an electric field we expect that non-trivial vevs might appear in the $A^{1,2}_{\mu}$ gauge fields. We therefore set
\be
\hat A_i=A^3_i=0\sp i=x,y
\label{PD5c} \ee
The remaining fields are $A^{1,2}_{t}$ and $A^{1,2}_{x,y}$.
As detailed in appendix \ref{AppA}, the bulk YM equations, expanded near the boundary \`a la FG imply that when $\m_3\not=0$ 
\be
A^{1,2}_t=0 \,.
\ee
We are therefore left with $A^{1,2}_{x,y}$.
So far the choice of sources has broken the global isospin $SU(2)$ symmetry to $SO(2)$ and we still have the $SO(2)$ rotation symmetry intact, that rotates the two spatial boundary coordinates $x,y$.

The remaining $SO(2)_\mathrm{isospin}\times SO(2)_\mathrm{rot}$ symmetry acts on the two-by-two matrix $A^i_{j}$ by an $SO(2)_\mathrm{rot}$ rotation on the left and an independent $SO(2)_\mathrm{isospin}$  rotation on the right
\be
\left(\begin{matrix} A^1_x & A^2_x\\ A^1_y &A^2_y\end{matrix}\right)\to \left(\begin{matrix} \cos\theta & \sin\theta\\ -\sin\theta&\cos\theta \end{matrix}\right)\cdot
\left(\begin{matrix} A^1_x & A^2_x\\ A^1_y &A^2_y\end{matrix}\right)
\cdot
\left(\begin{matrix} \cos\theta' & -\sin\theta'\\ \sin\theta'&\cos\theta' \end{matrix}\right)
\ee
The singular value theorem of linear algebra guarantees that we can always diagonalize this two-by-two matrix with independent left and right rotations.
Therefore, without loss of generality,
we are finally left with two non-zero order parameters $A^1_x$ and $A^{2}_{y}$.

We conclude that  the general ansatz, modulo symmetry is
\be
\label{PD5d} \mathbf{L} = \mathbf{R} = \frac{1}{2} \Phi(r) \intd t\,\mathbb{I}_2 + \frac{1}{2} \Phi_3(r) \intd t\,\s^3+ \frac{1}{2} A^1_x(r) \intd x\, \s^1 + \frac{1}{2} A^2_y(r) \intd y\, \s^2 \, .
\ee
Given the ansatz \eqref{PD5d}, the expected structure of the solutions is the following: they should form a discrete countable set, labeled by the number of nodes for each component of the condensed gauge field $(n,m)$, with $n$ the number of nodes\footnote{The point $r=0$ at the boundary is also counted as a node, such that the uncondensed (RN) solution is labeled $(0,0)$, and the condensed solutions start as $(0,1)$,$(1,0)$,$(1,1)$,\dots} for $A^1_x$, and $m$ for $A^2_y$. This set of solutions will obey additional properties:
\begin{itemize}
	
	\item By invariance of the background under spatial and isospin $SO(2)$ rotations, the solution $(n,m)$ is equivalent to the solution $(m,n)$. Therefore, the physically distinct solutions are labeled by $(n,m)$ with $n\leq m$;
	
	\item Since we expect\footnote{We verify this statement numerically later in this paper.} a single regular solution $(n,n)$ for each number $n$ of nodes, these solutions should be invariant under exchange of $A^1_x$ and $A^2_y$, and therefore have $A^1_x = A^2_y$. Those are the solutions that were computed in \cite{Gubser08a}, in the case of four bulk dimensions and vanishing baryon density. They preserve a $U(1)$ subgroup of the group $SO(2)_\mathrm{isospin}\times SO(2)_\mathrm{rot}$ of global symmetry and spatial rotations.
	
\end{itemize}

Among all the solutions, those that have nodes in the bulk (at $r>0$) are expected to be always subdominant, based on experience in holography, and our numerical results. This includes all levels $(n,m)$ with $n\leq m$ and $m\geq 2$. The dominant solution, that determines the phase of the theory, is among the following three remaining solutions
\begin{itemize}
	
	\item The solution labeled $(0,0)$ is the uncondensed solution, given by the Reissner-Nordstr\"om black hole in 3+1 dimensions, charged under both baryon number and isospin:
	\be
	\label{PD10} \mathrm{ds}^2 = \ex^{2A(r)}\left( -f(r)\mathrm{dt}^2 + f(r)^{-1}\mathrm{dr}^2 + \vec{\mathrm{d x}}^2 \right) \, ,
	\ee
	\be
	\label{PD11} \Phi(r) = \mu_q\left(1 - \frac{r}{r_H} \right) \sp \Phi_3(r) = \mu_3\left(1 - \frac{r}{r_H}\right) \, ,
	\ee
	where
	\be
	\label{PD12} \ex^{A(r)} = \frac{\ell}{r} \sp f(r) = 1 - \left(\frac{r}{r_H}\right)^3\left(1 + 3 - 4\pi T r_H \right) + \left(3 - 4\pi T r_H\right)\left(\frac{r}{r_H}\right)^4  \, ,
	\ee
	\be
	\label{PD13} r_H = \frac{3}{2\pi T} \left( 1 + \sqrt{1 + \frac{3w_0^2}{32N_c}\frac{\mu^2}{\pi^2T^2}} \right)^{-1} \, ,
	\ee
	and
	\be
	\label{PD14} \m \equiv \sqrt{\m_q^2+\m_3^2} \, .
	\ee
	
	\item The solution labeled $(0,1)$ is a solution where only one component of the gauge field condenses, that can be chosen to be $A^1_x$ without loss of generality. The corresponding ansatz for the gauge fields is given by
	\be
	\label{PD15} \mathbf{L} = \mathbf{R} = \frac{1}{2} \Phi(r) \intd t\,\mathbb{I}_2 + \frac{1}{2} \Phi_3(r) \intd t\,\s^3+ \frac{1}{2} A^1_x(r) \intd x\, \s^1 \, .
	\ee
	This configuration breaks the isospin $SO(2)$ to nothing, and the $SO(2)$ of spatial rotations to nothing. The appropriate ansatz for the metric is
	\be
	\label{PD16} \mathrm{ds}^2 = \ex^{2A(r)}\left( -f(r)\mathrm{dt}^2 + f(r)^{-1}\mathrm{dr}^2 + \intd x^2 + h(r)\intd y^2  \right) \, .
	\ee
	
	\item The solution labeled $(1,1)$ is a solution where two components of the gauge fields condense with the same amplitude, which can be written
	\be
	\label{PD17} \mathbf{L} = \mathbf{R} = \frac{1}{2} \Phi(r) \intd t\,\mathbb{I}_2 + \frac{1}{2} \Phi_3(r) \intd t\,\s^3+ \frac{1}{2} w(r) \big(\intd x\, \s^1 + \intd y\, \s^2\big) \, .
	\ee
	This configuration also breaks both the isospin $SO(2)$ and the spatial $SO(2)$, but preserves the axial subgroup of the isospin $SO(2)$ and the $SO(2)$. Consequently, the ansatz for the metric can be taken to be
	\be
	\label{PD18} \mathrm{ds}^2 = \ex^{2A(r)}\left( -f(r)\mathrm{dt}^2 + f(r)^{-1}\mathrm{dr}^2 + \intd x^2 + \intd y^2\right) \, .
	\ee
	
\end{itemize}

\subsection{$d=3+1$ and generalization to higher $d$}

The situation in higher number of spatial dimensions, including the case $d=3+1$ that interests us, is to a large extent similar to $d=2+1$, but the structure of the spatial rotation group changes.

The same arguments as for $d=2+1$ imply that the only components that may condense are $A^{1,2}_{i}$, with $i$ labeling the boundary spatial directions. The symmetry group preserved by the sources is now $SO(2)_{\text{isospin}}\times SO(d-1)_{\text{rot}}$, and the diagonalization argument discussed above also has a natural generalization.\footnote{Actually the argument also generalizes to higher number of flavors, one just needs to add additional columns in the matrix and consider chiral rotations of higher rank.} Namely, the rotation group acts on the matrix $A^i_{j}$ - which has two columns and $d-1$ rows - by an $SO(d-1)$ matrix $O_{d-1}$ on the left, whereas the isospin group acts with an $SO(2)$ rotation $P_2$ on the right:
\be
\left(\begin{matrix} A^1_x & A^2_x\\ A^1_y &A^2_y \\ \multicolumn{2}{c}{\vdots} \\ A^1_{(d-1)} &A^2_{(d-1)}\end{matrix}\right)\to O_{d-1}\cdot
\left(\begin{matrix} A^1_x & A^2_x\\ A^1_y &A^2_y \\ \multicolumn{2}{c}{\vdots}\\ A^1_{(d-1)} &A^2_{(d-1)}\end{matrix}\right)
\cdot
P_2 \,,
\ee
where the subscript $(d-1)$ stands for the last spatial index.
The singular value theorem guarantees that we can always diagonalize this matrix  to the form
\be
\left(\begin{matrix} ~~*~~ & ~~0~~\\ ~~0~~ & ~~*~~ \\ 0 &0\\ \multicolumn{2}{c}{\vdots}\\ 0 &0\end{matrix}\right)
\label{matr}\ee
where $*$ stands for a non-zero entry.
Therefore, as for $d=2+1$, there are two non-zero order parameters $A^1_x$ and $A^{2}_{y}$.

We conclude that the general gauge field ansatz takes the same form for any dimension $d$, namely
\be
\label{PD19} \mathbf{L} = \mathbf{R} = \frac{1}{2} \Phi(r) \intd t\,\mathbb{I}_2 + \frac{1}{2} \Phi_3(r) \intd t\,\s^3+ \frac{1}{2} A^1_x(r) \intd x\, \s^1 + \frac{1}{2} A^2_y(r) \intd y\, \s^2 \, ,
\ee
modulo symmetry. 

Given the ansatz \eqref{PD19}, the expected structure of the solutions, in terms of nodes is similar to that discussed in the previous subsection.
Among all the solutions, those that have nodes in the bulk (at $r>0$) are expected to be  subdominant. This includes levels $(n,m)$ with $n\leq m$ and $m\geq 2$. The dominant solution is among (0,0), (0,1) and (1,1), which for general $d$ are defined as
\begin{itemize}
	
	\item The solution $(0,0)$ is the uncondensed solution, given by the $(d+1)$-dimensional Reissner-Nordstr\"om black hole (see Appendix~\ref{Sec:bgd}), charged under both baryon number and isospin:
	\be
	\label{PD20} \mathrm{ds}^2 = \ex^{2A(r)}\left( -f(r)\mathrm{dt}^2 + f(r)^{-1}\mathrm{dr}^2 + \vec{\mathrm{d x}}^2 \right) \, ,
	\ee
	\be
	\label{PD21} \Phi(r) = \mu_q\left(1 - \left(\frac{r}{r_H}\right)^{d-2}\right) \sp \Phi_3(r) = \mu_3\left(1 - \left(\frac{r}{r_H}\right)^{d-2}\right) \, ,
	\ee
	where
	\be
	\label{PD22} \ex^{A(r)} = \frac{\ell}{r} \sp f(r) = 1 - \left(\frac{r}{r_H}\right)^d
	+ \frac{d - 4\pi T r_H}{d-2}\left[\left(\frac{r}{r_H}\right)^{2d-2}-\left(\frac{r}{r_H}\right)^d\right]  \, ,
	\ee
	\be
	\label{PD23} r_H = \frac{d}{2\pi T} \left( 1 + \sqrt{1 + \frac{d(d-2)^2}{16(d-1)}\frac{w_0^2}{N_c}\frac{\mu^2}{\pi^2T^2}} \right)^{-1} \, ,
	\ee
	and
	\be
	\label{PD24} \m \equiv \sqrt{\m_q^2+\m_3^2} \, .
	\ee
	
	\item The solution labeled $(0,1)$ is a solution where only one component of the gauge field condenses. We adopt a naming convention where we call the corresponding spatial coordinate $z$, and the other coordinates $x_1, x_2, \ldots , x_{d-2}$.   Therefore the condensing component is $A^1_z$, and the ansatz for the gauge fields is given by
	\be
	\label{PD25} \mathbf{L} = \mathbf{R} = \frac{1}{2} \Phi(r) \intd t\,\mathbb{I}_2 + \frac{1}{2} \Phi_3(r) \intd t\,\s^3+ \frac{1}{2} A^1_z(r) \intd z\, \s^1 \, .
	\ee
	This configuration breaks the isospin $SO(2)$ to nothing, and the $SO(d-1)$ of spatial rotations to the $SO(d-2)$ subgroup that keeps the $z$-component unchanged and rotates the coordinates $x_1, x_2, \ldots , x_{d-2}$.
	With this symmetry, the appropriate ansatz for the metric is
	\be
	\label{PD26} \mathrm{ds}^2 = \ex^{2A(r)}\left( -f(r)\mathrm{dt}^2 + f(r)^{-1}\mathrm{dr}^2 + \intd x_1^2 + \cdots + \intd x_{d-2}^2  + h(r) \intd z^2 \right) \, .
	\ee
	
	\item The solution labeled $(1,1)$ is a solution where two components of the gauge fields condense with the same amplitude, which can be written
	\be
	\label{PD27} \mathbf{L} = \mathbf{R} = \frac{1}{2} \Phi(r) \intd t\,\mathbb{I}_2 + \frac{1}{2} \Phi_3(r) \intd t\,\s^3+ \frac{1}{2} w(r) \big(\intd x\, \s^1 + \intd y\, \s^2\big) \, .
	\ee
	This configuration also breaks both the isospin $SO(2)$ and the spatial $SO(d-1)$, but preserves the axial subgroup of the isospin $SO(2)$ and the $SO(2)$ that rotates the $(x,y)$ plane, as evident from (\ref{matr}). It also leaves unbroken the $SO(d-3)$ rotations of the remaining spatial coordinates which we denote as $z_1, z_2, \ldots , z_{d-3}$. Consequently, the ansatz for the metric can be written as
	\be
	\label{PD28} \mathrm{ds}^2 = \ex^{2A(r)}\left( -f(r)\mathrm{dt}^2 + f(r)^{-1}\mathrm{dr}^2 +  \intd x^2 + \intd y^2 + h(r)\left(\intd z_1^2 +  \cdots + \intd z_{d-3}^2\right)  \right) \, .
	\ee
	Note that in the special case $d=3+1$ that we  discuss below, this ansatz coincides with that of the $(0,1)$ solution \eqref{PD26} after renaming the spatial coordinates.
	
\end{itemize}

The equations of motion for each solution $(0,1)$ and $(1,1)$ are listed in Appendix \ref{AppB}.

\section{The phase diagram}

\label{Sec:PD}

We discuss in this section the phase diagram for the model \eqref{Sb}. We start by describing the method in general boundary dimension $d\geq 3$, before presenting our results for the cases of $d=2+1$ and $d=3+1$. Even though no qualitative difference is observed between the two cases, starting from $d=2+1$ makes an explicit connection with existing results in \cite{Gubser08a,Gubser08b}.

In order to construct the phase diagram, we  need to identify the leading solution for given (inverse) flavor coupling $w_0$. Since we have in mind a physical system with a fixed number of particles such as a neutron star, we choose to present the results in the canonical ensemble\footnote{At the level of the phase diagram, the canonical and grand canonical ensembles will differ where the phase transitions are first order. Our numerical results indicate that there is no qualitative difference between the two.}. Hence, for given isospin and quark number density $n_3$ and $n_q$, we  look for the solution with the smallest (Helmholtz) free energy. According to the discussion of the previous section, we need to consider three competing solutions: the uncondensed Reissner-Nordstr\"om solution \eqref{PD10}-\eqref{PD13}, the (0,1) condensate \eqref{PD16}, and the (1,1) condensate \eqref{PD18}.

The Reissner-Nordstr\"om solution is known analytically, and given by the expressions \eqref{PD10}-\eqref{PD13}. In contrast, the condensed solutions obey the equations of motion written in appendix \ref{AppB}, which can only be solved numerically. We use for this a standard shooting method, where regular boundary conditions are imposed at the horizons, and physical observables are extracted from the near-boundary data. At finite temperature, the regular behavior of the fields near the horizon $(r\to r_H)$ is given by :
\begin{equation}
\label{d30a} \Phi(r) = \Phi'(r_H)(r-r_H)  + \OO(r-r_H)^2 \sp \Phi_3(r_H) = \Phi_3'(r_H)(r-r_H)  + \OO(r-r_H)^2 \, ,
\end{equation}
\begin{equation}
\label{d30b} A(r) = A(r_H) + \OO(r-r_H) \sp f(r) = f'(r_H)(r-r_H) + \OO(r-r_H)^2\, ,
\end{equation}
\begin{equation}
\label{d30c} (0,1) \,:\, h(r) = h(r_H) + \OO(r-r_H) \sp A^1_z(r) = A^1_z(r_H) + \OO(r-r_H) \, ,
\end{equation}
\begin{equation}
\label{d30d} (1,1) \,:\, w(r) = w(r_H) + \OO(r-r_H) \, .
\end{equation}
The leading coefficients in the near horizon expansions above give the free IR parameters, that fully determine all the higher order coefficients. Among those, the parameters related to the metric $A(r_H),f'(r_H)$ and $h(r_H)$, correspond to the residual freedom of coordinates rescalings. The last two are fixed by imposing that $f(r)$ and $h(r)$ go to 1 at the boundary. Then, the remaining choice of $A(r_H)$ defines the unit in which the dimensionful quantities are measured. We choose to set the temperature $T = -f'(r_H)/(4\pi)$ to 1, such that all quantities are expressed in units of the temperature.

The coefficients for the gauge fields $\Phi'(r_H),\Phi_3'(r_H)$ and $A^1_z(r_H)$ (or $w(r_H)$) correspond to the actual (dimensionless) parameters of the theory. That is, they give an alternative parametrization of the space of boundary sources, which appear in the leading near-boundary expansion of the gauge fields
\begin{equation}
\nn \Phi(r) = \mu_q  + \OO(r^{d-2}) \sp \Phi_3(r) = \mu_3 + \OO(r^{d-2},r^2) \, ,
\end{equation}
\begin{equation}
\label{d31} (0,1) \,:\, A^1_z(r) = A^{1,(0)}_z  + \OO(r^{d-2},r^2)\sp (1,1) \,:\, w(r) = w^{(0)} + \OO(r^{d-2},r^2)\, .
\end{equation}
$\mu_q$ and $\mu_3$ are the quark number and isospin chemical potentials, whereas $A^{1,(0)}_z$ (or $w^{(0)}$) gives a source to the condensing current. For generic IR parameters, this source will not be zero. Since we are looking for configurations where the gauge field is not sourced but condenses spontaneously, we need to scan over the IR parameters until we reach a solution where the source vanishes. This can be done efficiently with a bisection method.

Once the sources are set to zero for the condensing gauge fields, the remaining sources and dual operator expectation values appear in the near-boundary ($r\to 0$) expansion of the bulk fields as
\begin{equation}
\nn \Phi(r) = \mu_q + \Phi^{(d-2)} r^{d-2}(1+\OO(r^d)) \sp \Phi_3(r) = \mu_3 + \Phi_3^{(d-2)} r^{d-2}(1+\OO(r^d)) \, ,
\end{equation}
\begin{equation}
\label{d32} A(r) = \log{\left(\frac{\ell}{r}\right)} + A^{(d)} r^d(1+\OO(r^{d-2})) \sp f(r) = 1 + f^{(d)} r^d(1+\OO(r^{d-2})) \, ,
\end{equation}
\begin{equation}
\nn h(r) = 1 + h^{(d)} r^d(1+\OO(r^{d-2})) \, ,
\end{equation}
\begin{equation}
\nn (0,1) \,:\, A^1_z(r) = A^{1,(d-2)}_z r^{d-2}(1+\OO(r^2)) \sp (1,1) \,:\, w(r) = w^{(d-2)} r^{d-2}(1+\OO(r^2)) \, .
\end{equation}
The precise relation to the dual operators one-point functions can be derived by substituting these expansions in the on-shell action\footnote{The relation between the near-boundary expansion of the fields and the dual operator expectation values is straightforward for the currents, but more subtle for the stress tensor, since the on-shell action needs to be renormalized. Fortunately, the expression that results from this procedure is available in \cite{deHaro00}. Their result is in principle modified with scheme-dependent terms in presence of gauge-fields, but those terms vanish for zero field strengths at the boundary.} \eqref{Sb}. For the densities and quark-pair condensates we obtain
\begin{equation}
\nn n_q \equiv \frac{1}{2}\left<\hat{J}_L^t + \hat{J}_R^t\right> = -(M\ell)^{d-1}\frac{(d-2)w_0^2}{2N_c}  \Phi^{(d-2)} \, ,
\end{equation}
\begin{equation}
\label{d33} n_3 \equiv \frac{1}{2}\left<J_L^{3,t} + J_R^{3,t}\right> = -(M\ell)^{d-1}\frac{(d-2)w_0^2}{2N_c}  \Phi_3^{(d-2)} \, ,
\end{equation}
\begin{equation}
\nn  (0,1) \,:\, J_z \equiv \frac{1}{2}\left<J_L^{1,z} + J_R^{1,z}\right>  = (M\ell)^{d-1}\frac{(d-2)w_0^2}{2N_c}  A_z^{1,(d-2)} \, ,
\end{equation}
\begin{equation}
\label{d34} (1,1) \,:\, J_{xy} \equiv \frac{1}{2}\left<J_L^{1,x} + J_R^{1,x}\right> = \frac{1}{2}\left<J_L^{2,y} + J_R^{2,y}\right> = (M\ell)^{d-1}\frac{(d-2)w_0^2}{N_c}  w^{(d-2)} \, ,
\end{equation}
where the decomposition of the chiral currents in the SU(2) basis is as in \eqref{JsP}. The stress tensor expectation value is extracted from\footnote{We used the constraint from conformality $\left<T^\mu_\mu\right> = 0$, to remove $A^{(d)}$ from the expressions. As usual, this condition is reproduced holographically from the near-boundary expansion of the constraint from the Einstein equations \eqref{B5}.}
\begin{equation}
\label{d35} (0,1) \,:\,
\begin{array}{ c }
\EE \equiv \left<T_{00}\right> = - (M\ell)^{d-1}N_c^2 ((d-1) f^{(d)} - h^{(d)}) \, ,\\
p_x \equiv \left<T_{x_1x_1}\right> = \dots = \left<T_{x_{d-2}x_{d-2}}\right> = - (M\ell)^{d-1}N_c^2 (f^{(d)} + h^{(d)}) \, , \\
p_z \equiv \left<T_{zz}\right> = - (M\ell)^{d-1}N_c^2 (f^{(d)} - (d-1) h^{(d)}) \, ,
\end{array}
\end{equation}
\begin{equation}
\label{d36} (1,1) \,:\,
\begin{array}{ c }
\EE \equiv \left<T_{00}\right> = - (M\ell)^{d-1}N_c^2 ((d-1) f^{(d)} - (d-3)h^{(d)}) \, ,\\
p_x \equiv \left<T_{xx}\right> = \left<T_{yy}\right> = - (M\ell)^{d-1}N_c^2 (f^{(d)} + (d-3)h^{(d)}) \, , \\
p_z \equiv \left<T_{z_1z_1}\right> = \dots = \left<T_{z_{d-3}z_{d-3}}\right> = - (M\ell)^{d-1}N_c^2 (f^{(d)} - 3 h^{(d)}) \, .
\end{array}
\end{equation}
We denoted by $T_{\m\n}$ the boundary stress tensor, by $\EE$ the boundary energy density, and by $p_{x_i}$ the pressure in the direction of $x_i$. The free energy density is related to the energy by
\begin{equation}
\label{d36b} \FF = \EE - T s \, ,
\end{equation}
with the entropy density $s$ simply computed from the area of the horizon
\begin{equation}
\label{d37} s = 4\pi (M\ell)^{d-1}N_c^2\, \ex^{(d-1)A(r_H)}h(r_H)^{\frac{a}{2}} \, ,
\end{equation}
where $a = 1$ for (0,1), and $d-3$ for (1,1).

The paragraphs above give all the ingredients to compute numerically the bulk solutions corresponding to the different phases of the theory, and to extract the relevant observables. In particular, \eqref{d35}-\eqref{d37} can be used to compare the free energies of each phase, and infer the phase diagram. The next subsections present the results of this numerical analysis, first for $d=2+1$ and then for $d=3+1$.

\subsection{$d=2+1$}

\label{Sec:PD2d}

We discuss in this subsection the phase diagram for the model \eqref{Sb} in the case of $d=2+1$ boundary dimensions. This problem was partially addressed in \cite{Gubser08a}, where it was shown that the (1,1) condensate \eqref{PD17} dominates over the uncondensed phase below a certain critical temperature $(T/\mu_3)_c(w_0)$, where a second order phase transition happens. Here, we  complete this analysis by considering the most general ansatz for the condensate \eqref{PD5d}. That is, we include the possibility that the (0,1) condensate dominates over (1,1), which will turn out to be the case. We shall not introduce a baryon number chemical potential in this section, since our focus will be on the comparison with \cite{Gubser08a}. The dependence on baryon number is expected to be essentially the same as in 3+1 boundary dimensions, which is discussed in the next subsections.

\begin{figure}[h!]
	\begin{center}
		\includegraphics[scale=0.85]{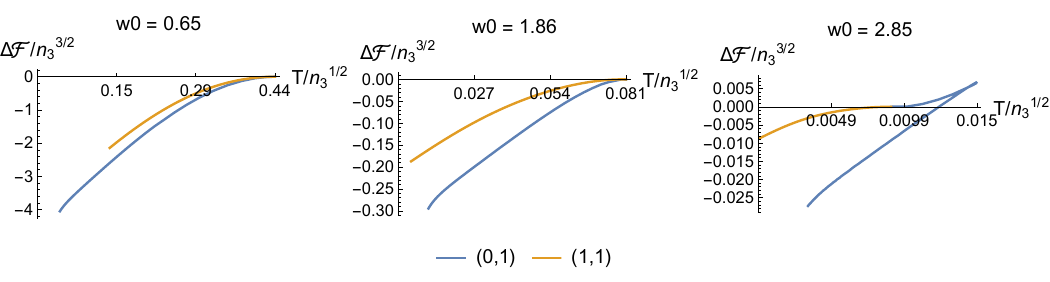}
		\caption{For $d=2+1$, free energy difference of the paired phases ((0,1) in blue and (1,1) in orange) with the uncondensed Reissner-Nordstr\"om solution, as a function of temperature, and for three different values of $w_0$. All quantities are expressed in units of the isospin density.}
		\label{fig:Dfw03d}
	\end{center}
\end{figure}

\begin{figure}[h!]
	\begin{center}
		\includegraphics[scale=1.]{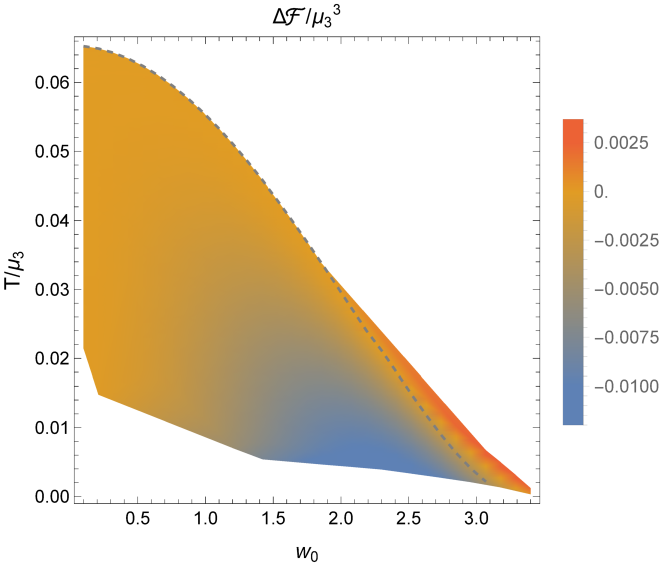}
		\caption{Free energy difference of the (0,1) phase with the (1,1) phase, over the range of temperature and $w_0$ where the solutions exist. We found it convenient for visualizing the results to express all quantities in units of the isospin chemical potential in the leading (0,1) solution (which is not the same as in RN and (1,1) since we consider the canonical ensemble). The gray dashed line indicates the temperature beyond which the (1,1) solution ceases to exist. Beyond that line, what is shown is the free energy difference of (0,1) with the RN solution.}
		\label{fig:Dffull3d}
	\end{center}
\end{figure}

Figure \ref{fig:Dfw03d} shows the free energy difference of the (0,1) and (1,1) condensed phases with the unbroken Reissner-Nordstr\"om phase, for three different values of the coupling $w_0$. Note that because constructing solutions numerically at low temperatures is challenging, our curves end at nonzero value of the temperature, even though the solutions exist all the way down to zero temperature.
These results indicate that on the one hand, (0,1) seems to always dominate over (1,1), and on the other hand the transition becomes first order at large enough $w_0$. Those claims are verified by calculating the free energy difference for all values of $w_0$, as shown in figure \ref{fig:Dffull3d}.

As is clear from figure \ref{fig:Dfw03d} right, two subtleties have to be taken into account when constructing figure \ref{fig:Dffull3d} in the region where the transition becomes first order. First, at high enough temperatures, the free energy is multivalued in the (0,1) phase, with two branches of solutions. The branch that we show in figure \ref{fig:Dffull3d} is the stable one, corresponding to the lower branch in figure \ref{fig:Dfw03d}. Second, precisely at the temperature where the (0,1) free energy becomes multivalued, the (1,1) solution ceases to exist. In that region, delimited by the gray dashed line in figure \ref{fig:Dffull3d}, what we show is the free energy difference of (0,1) with Reissner-Nordstr\"om. Those conventions make it possible to show in the same plot the dominance of the (0,1) phase over (1,1), and the onset of a first order transition at large $w_0$. Note that we are again leaving out the low temperature region of the plot because contructing the free energy difference numerically at low temperatures is difficult.

\begin{figure}[h!]
	\begin{center}
		\includegraphics[scale=0.9]{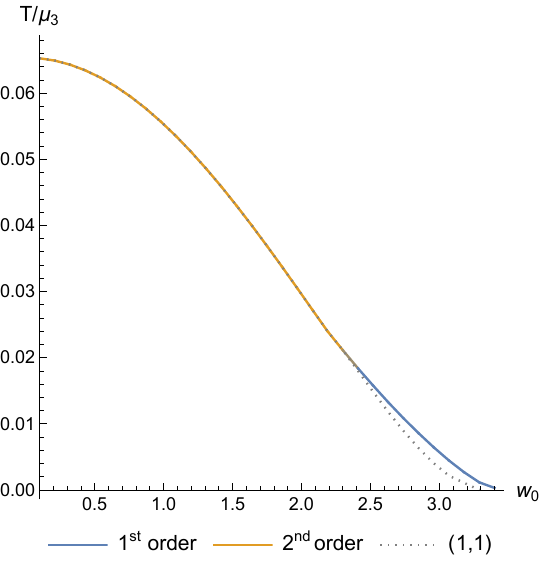}
		\caption{Phase transition line for $d=2+1$, with the Reissner-Nordstr\"om solution above, and the (0,1) paired phase below. The transition is second order at low $w_0$ (in orange), and first order at large $w_0$ (in blue). For comparison, we indicate with the gray dotted line the second order transition line computed in \cite{Gubser08a}, based on the (1,1) solution. As in figure \ref{fig:Dffull3d}, we expressed the temperature in units of the chemical potential in the (0,1) solution.}
		\label{fig:PTl3d}
	\end{center}
\end{figure}

Based on the results for the free energy, the phase diagram of the theory can be constructed in the plane of $w_0$ and temperature, as shown in figure \ref{fig:PTl3d}. For comparison, we also include the result obtained in \cite{Gubser08a} by considering only the (1,1) condensate\footnote{Our definitions differ from \cite{Gubser08a}. To compare the two results, we use that the parameter $w_0$ is related to the flavor Yang-Mills coupling $g$ by $w_0 = 2\sqrt{N_c/N_f}g^{-1}$. Also, the normalization of the gauge fields is different : $\mathbf{A}_\mu = g\tilde{\mathbf{A}}_\mu$, with a tilde referring to the definition of \cite{Gubser08a}.}. The main difference with \cite{Gubser08a} is that the leading superconducting solution is the (0,1) condensate, which was not considered in \cite{Gubser08a}. This result is consistent with \cite{Gubser08b}, where the (0,1) solution was considered in the probe limit $w_0\to 0$, and found to be perturbatively stable, while (1,1) was found to feature an instability. As far as the phase transition is concerned, we observe that the two transition lines agree when it is second order, whereas they start differing when the transition to the (0,1) phase becomes first order. This implies in turn that the symmetry broken phase covers a slightly larger part of the phase space compared with \cite{Gubser08a}.

\begin{figure}[h!]
	\begin{center}
		\includegraphics[scale=0.85]{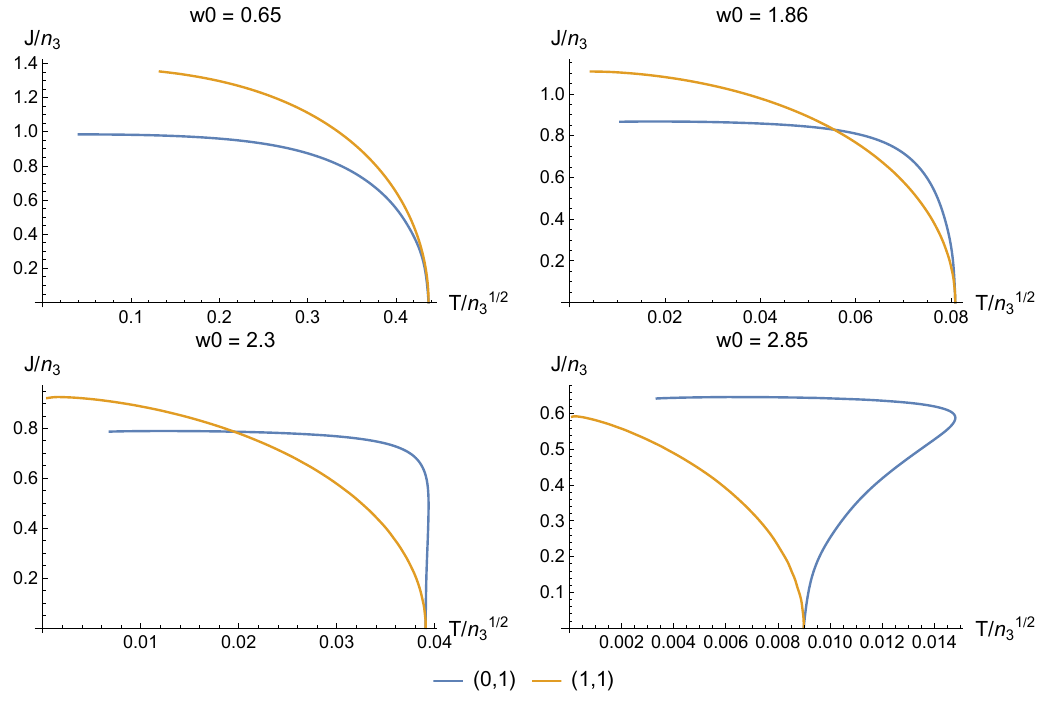}
		\caption{For $d=2+1$, order parameter \eqref{d34} for the (0,1) (blue) and (1,1) (orange) paired phases as a function of temperature, and for four different values of $w_0$. All quantities are expressed in units of the isospin density.}
		\label{fig:OPw0}
	\end{center}
\end{figure}

Even though the phase diagram in figure \ref{fig:PTl3d} does not show any difference between the (0,1) and (1,1) phases in the second order region, a closer inspection of the transition for the two cases shows a different  behavior as $w_0$ is increased. Figure \ref{fig:OPw0} shows the order parameter of the transition as a function of temperature, for several values of $w_0$, and for the two types of pairing (0,1) and (1,1). The (0,1) transition is observed to become sharper and sharper as $w_0$ is increased, until it turns first order at $w_{0,1} \simeq 2.2$. In terms of critical exponents, this means that the exponents decrease\footnote{A closer inspection of the numerical results indicates that this decrease is mostly concentrated in a very narrow region close to $w_{0,1}$.} with $w_0$, from the mean field values at $w_0 = 0$ (in particular $J_{z,xy}\sim (T_c-T)^{1/2}$ as $T\to T_c$), to zero at $w_0 = w_{0,1}$. As for the (1,1) transition, no qualitative change is observed as $w_0$ is increased, and the critical exponents remain equal to those predicted by mean-field theory.

In the next subsections, we analyze the main case of interest for us, corresponding to $d=3+1$ boundary dimensions. As we shall see, the structure of the phase diagram in the plane of $w_0$ and temperature will be essentially the same, but the analysis will be pushed further in two directions. First, we consider introducing finite baryon density, and second, we  give a complete analysis of the condensed solution at zero temperature.

\subsection{$d=3+1$ at finite temperature}

\label{Sec:PD3d}

We analyze in this subsection the case of $d=3+1$ boundary dimensions. We still follow the general method outlined at the beginning of this section, but now including a finite baryon density. For now we consider general finite temperature, leaving the special case of zero temperature to the next subsection. The numerical results are first presented, and it is then discussed which parts of the phase diagram are expected to be relevant to a neutron-star-like environment.

\begin{figure}[h!]
	\begin{center}
		\includegraphics[scale=1.]{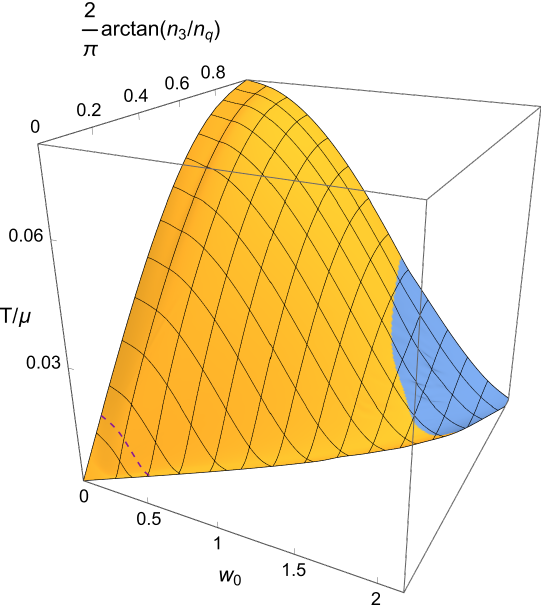}
		\caption{Phase transition surface for $d=3+1$, between the Reissner-Nordstr\"om solution above, and the (0,1) paired phase below. The orange part is second order, whereas the blue part is first order. The phase space is spanned by $w_0$, the ratio of isospin to quark number density $n_3/n_q$, and the temperature measured in units of the chemical potential $\m = \sqrt{\m_q^2+\m_3^2}$. When the transition is first order, $\m$ is the chemical potential in the (0,1) phase. We use the compact coordinate $\arctan(n_3/n_q)$ in order to show both the planes at $n_3 = 0$ and $n_q = 0$. The purple dashed line indicates the maximal value of $n_3/n_q$ compatible with charge neutrality and $\b$-equilibrium (see appendix \ref{CN}).}
		\label{fig:PDnq4d}
	\end{center}
\end{figure}

\begin{figure}[h!]
	\begin{center}
		\includegraphics[scale=1.]{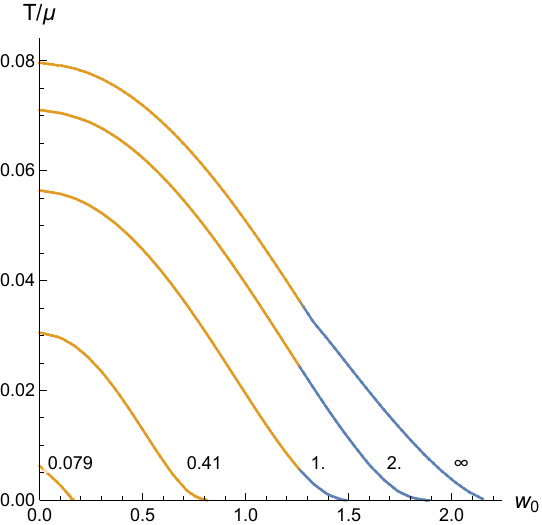}
		\caption{Cuts of figure \ref{fig:PDnq4d} at constant values of $n_3/n_q$, indicated at the bottom of each line.}
		\label{fig:PDnq4d_rqx}
	\end{center}
\end{figure}

\begin{figure}[h!]
	\begin{center}
		\hspace{-2cm}
		\includegraphics[scale=1.]{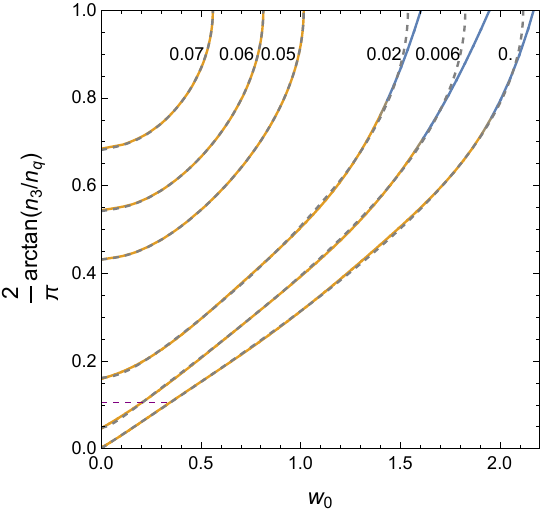}
		\caption{Cuts of figure \ref{fig:PDnq4d} at constant values of $T/\m$, indicated at the top of each line. The gray-dashed lines show the exact result for the dependence on the ratio of densities in the second order region \eqref{d41}. The purple dashed line is the projection of the line shown in figure \ref{fig:PDnq4d}.}
		\label{fig:PDnq4d_Tx}
	\end{center}
\end{figure}

\begin{figure}[h!]
	\begin{center}
		\hspace{1cm}
		\includegraphics[scale=1.2]{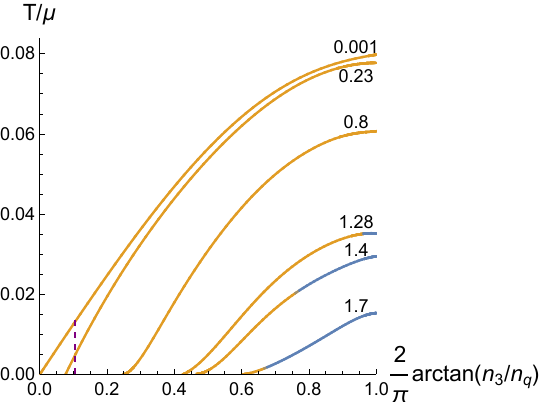}
		\caption{Cuts of figure \ref{fig:PDnq4d} at constant values of $w_0$, indicated above each line. The purple dashed line is the projection of the line shown in figure \ref{fig:PDnq4d}.}
		\label{fig:PDnq4d_w0x}
	\end{center}
\end{figure}

Figure \ref{fig:PDnq4d} shows the full three-dimensional phase diagram as a function of $w_0$, the temperature, and the ratio of isospin to quark number density $n_3/n_q$, with two-dimensional cuts presented in figures \ref{fig:PDnq4d_rqx} to \ref{fig:PDnq4d_w0x}. As in $d=2+1$, the model enters a paired phase below a certain critical temperature $T_c(w_0,n_3/n_q)$, where it is always the $(0,1)$ condensate which dominates.

From figure \ref{fig:PDnq4d_rqx}, we see that the qualitative behavior in the $(w_0,T)$ plane is similar to $d=2+1$ (figure \ref{fig:PTl3d}). For large values of $n_3/n_q$, the transition is second order at low $w_0$, becomes first order when $w_0$ is increased beyond a certain value $w_{0,1}(n_3/n_q)$, until the paired phase completely disappears at a finite value $w_{0,c}(n_3/n_q)$. For smaller $n_3/n_q$, the interval with first order transition is absent.
The critical value $w_{0,c}$ increases monotonically from $w_{0,c}(0) = 0$ to $w_{0,c}(\infty) \simeq 2.18$, the former result being consistent with the known absence of a superconducting phase at zero isospin density.

The full dependence of $w_{0,c}$ on $n_3/n_q$ is shown in figure \ref{fig:PDnq4d_Tx} (the $T = 0$ line), together with other constant-temperature lines of the phase transition surface. The latter are observed to behave similarly to the zero-temperature line. As indicated in figure \ref{fig:PDnq4d_Tx}, it is actually possible to derive an analytic expression for the density dependence of the phase transition surface, in the region where it is second order. The equation for the transition lines at constant $T/\mu$ is of the form
\begin{equation}
\label{d41} w_0 \left(1 + \frac{n_q^2}{n_3^2}\right)^{\frac{1}{2}} = w_0^*\left(\frac{T}{\mu_3}\right) = w_0^*\left(\frac{T}{\mu}\left(1 + \frac{n_q^2}{n_3^2}\right)^{\frac{1}{2}}\right) \, ,
\end{equation}
where $w_0^*$ is a function that can be found numerically.
The simple argument that leads to this result is presented in appendix \ref{NBpt}, where it is also explained what is the numerical data required to compute the function $w_0^*$.

The phase diagram of the model that was analyzed shows an interesting structure, with a low temperature superconducting phase at finite isospin density. The latter may have very different transport properties from the RN phase, and be of relevance in particular for neutrino transport in neutron star-like conditions, that was analyzed for this model at $n_3 = 0$ in \cite{neutrino}. However, whether the condensed phase is relevant or not also depends on physical constraints on the phase space. In particular, the ratio of isospin to quark number density is bounded for a neutral medium at $\b$-equilibrium (see appendix \ref{CN} for details)
\begin{equation}
\label{d42} \frac{|n_3|}{n_q} \in \left[0,\frac{1}{6}\right] \, .
\end{equation}
Furthermore, the flavor parameter $w_0$ cannot be arbitrarily small or large for a realistic model, which comes both from top-down and bottom-up arguments. On the one hand, in string theoretic constructions, a Yang-Mills action such as \eqref{SYM} typically arises as the expansion at lowest order in derivatives of the DBI action controlling the dynamics of flavor branes \cite{Karch02,Sakai03}. Assuming that \eqref{SYM} comes from flavor branes\footnote{In doing the comparison with flavor branes, we assume that the solution with the branes does not involve more fields that we considered here, i.e. the metric and gauge fields. In particular, we ignore the dynamics of the dilaton, and the scalar field dual to the chiral condensate. Those fields are known to have non-trivial profiles in actual calculations with flavor branes \cite{Karch02,Sakai03,Nunez10}, so that our set-up does not have a simple top-down completion. However, we still expect that the value of $w_0$ from the DBI action should give the typical order of magnitude predicted by string theory.}
would imply that $w_0$ is equal to 2.

On the other hand, the flavor dependence of the pressure is proportional to $w_0^2$, and should not be too far from actual QCD if the model is to give at least some suggestive ideas for the behavior of neutron star matter. In \cite{neutrino}, it was shown that fitting the model parameters to the quark gluon plasma (QGP) thermodynamics required $w_0 \simeq 3.72$.

The bound on isospin density \eqref{d42} is shown in figures \ref{fig:PDnq4d}, \ref{fig:PDnq4d_Tx} and \ref{fig:PDnq4d_w0x}, as the purple-dashed line. The constraint is observed to confine the condensed phase to the region of very low temperature and $w_0$. Together with the arguments above that $w_0$ should be of order 1 in a realistic set-up, this indicates that the superconducting phase is unlikely to be relevant to physically relevant conditions, including neutron star matter. We should recall however that the model used here is not very close to real QCD, and a more solid statement would require to repeat our analysis in a more realistic model of holographic QCD.

\subsection{$d=3+1$ at zero temperature}

\label{Sec:T0}

The bulk solutions that correspond to each point in the phase diagram (figure \ref{fig:PDnq4d}) are qualitatively the same for every generic values of the parameters in a given phase, with Reissner-Nordstr\"om above the transition surface, and hairy black holes below. However, new kinds of solutions typically arise in the limit of vanishing temperature \cite{Horowitz09,Charmousis:2010zz,Gouteraux12}, that correspond to the zero temperature phases of the dual field theory. Those solutions are largely characterized by the IR geometry, that determines the effective theory at low energy. Another important property is the distribution of charges in the bulk, with charges hidden behind an (extremal) horizon corresponding to a \emph{fractionalized} (or deconfined) phase, whereas charged degrees of freedom outside the horizon realize a \emph{cohesive} (or confined) state \cite{Gouteraux12}.

The uncondensed phase is given by the  
extremal RN solution, but the condensed superconducting solution is a priori unknown. Even though it can be guessed from the behavior of the numerical hairy black hole solution at low temperature, its exact (numerical) calculation requires to identify the proper IR boundary conditions in this limit. This constitutes the content of this subsection, where the numerical solution at zero temperature is also presented. We focus on the leading (0,1) solution, and start by discussing generic values of $n_3/n_q$, before analyzing the specific case $n_q = 0$.

\subsubsection{Finite quark number density}

\label{T0nq}

Since the color sector of our model is $\NN=4$ SYM, which is not confining, cohesive states charged under quark number (baryons) are not expected to exist\footnote{Actually, since we do not include a Chern-Simons term in this work, the cohesive baryon number will be zero for any configuration (see e.g. \cite{SS}). Our argument here implies that this should still hold for every regular bulk solution when the Chern-Simons is included.}. Hence, at finite quark density $n_q$, the geometry in the deep IR should contain an extremal horizon with a quark number charge proportional to $n_q$. The horizon will lie at a finite value of the holographic coordinate $r_H$, where the metric should behave as
\begin{equation}
\label{T0n1} f(u) \sim  \frac{1}{2}f''(0) u^2 \sp A(u) \sim A(0) \sp h(u) \sim h(0) \sp u \equiv \frac{r_H - r}{r_H} \, .
\end{equation}
As for the gauge fields, the time components $\Phi$ and $\Phi_3$ should go to zero at $u=0$ for regularity, whereas $A^1_z$ may go to a finite value $A^1_z(0)$. The leading behavior of $\Phi$ and $\Phi_3$ as $u$ goes to 0, as well as the sub-leading behavior of the other fields, are determined by the near horizon limit of the equations of motion \eqref{B3}-\eqref{B8}. In particular, the temporal components are found to behave as a power-law near the horizon
\begin{equation}
\label{T0n2} \Phi(u) \sim \Phi'(0) u \sp \Phi_3(u) \sim \Phi_{3,0}u^{c_3}\, , \,\,\, c_3 > 1 \, .
\end{equation}

At higher order in the near-horizon expansion, the fields can be expanded as a series involving generically non-integer powers of $u$. Since the only parameters characterizing the state at the boundary are $n_3/n_q$ and the source for the order parameter $A^{1,(0)}_z/n_3^{1/3}$, all the coefficients and exponents in the series will be determined in terms of two parameters\footnote{As usual, the leading order IR coefficients for the metric in \eqref{T0n1} are also free, which corresponds to trivial rescalings of the boundary theory.}. One of those will be fixed by requiring that the source $A^{1,(0)}_z$ vanishes, and we are left with one IR degree of freedom, for which a convenient choice is the exponent $c_3$. The fields are then found to obey an expansion as a double series, in $u$ to some $c_3$-dependent powers $\a(c_3)$ and $\g(c_3)$
\begin{equation}
\label{T0n3} \f(u) = u^{c_\f}\sum_{j,k \in \mathbb{N}} \f_{jk}(c_3) u^{j\, \a(c_3)} u^{k\, \g(c_3)} \sp \f\in\{A,f,h,\Phi,\Phi_3,A^1_z\} \, ,
\end{equation}
where $c_\f$ is the leading power-law for each field : $c_\f = 0$ for $A$,$h$ and $A^1_z$, $c_f = 2$, $c_{\Phi} = 1$ and $c_{\Phi_3} = c_3$.

The expansion \eqref{T0n3} takes a different form depending on whether $c_3$ is smaller or larger than 3/2. In particular, the exponents $\a(c_3)$ and $\g(c_3)$ are given by
\begin{equation}
\label{T0n4} \a(c_3) =
\left\{
\begin{array}{rl}
2c_3 - 2 & \, ,\,\, 1< c_3 < \frac{3}{2} \\
1 & \, ,\,\, c_3 > \frac{3}{2}
\end{array}
\right. \sp \g(c_3) = \left|2c_3 - 3\right| \, .
\end{equation}
Note that the expansion breaks down at $c_3 = 3/2$, where $\g(c_3)$ goes to 0. In this limit, the expansion \eqref{T0n3} involves log terms $\log{u}^{-j}$ rather than power-laws. Another peculiar point is at $c_3 = 1$, where $\a(c_3)$ has a zero. The case $c_3 = 1$ includes extremal RN solutions, and it is easy to check that there is no other solution in this limit.

The coefficients $\f_{jk}$ that determine the IR behavior of the fields, can be computed by solving the equations of motion \eqref{B3}-\eqref{B8} order by order in the double expansion \eqref{T0n3}. Up to first order in $u^{\g(c_3)}$, the corrections are given by
\begin{equation}
\label{T0n5} A(u) =
\left\{
\begin{array}{ll}
A_{0} + v^{2c_3-2}\left(1 - 2(2-c_3)\frac{w_0^2 A^1_{z,1}\ell}{N_c} v^{3-2c_3} +  \OO(v^{2c_3-2},v^{2(3-2c_3)})\right) &, 1< c_3 < \frac{3}{2} \\
A_{0} + v\left(1 - \frac{c_3(c_3-1)^2}{3(2c_3-3)^2(2c_3-1)}\frac{w_0^2\Phi_{3,0}^2\ell^2}{N_c}v^{2c_3 - 3} + \OO(v, v^{2(2c_3 - 3)}) \right) &, c_3 > \frac{3}{2}
\end{array}
\right.
\end{equation}
\begin{equation}
\label{T0n6} f(u) =
\left\{
\begin{array}{ll}
f_{0}v^2\left(1 + \frac{w_0^2A^1_{z,1}\ell}{N_c(c_3-1)} v^{3-2c_3} + \OO(v^{2c_3-2},v^{2(3-2c_3)}) \right)  &, 1< c_3 < \frac{3}{2} \\
f_{0}v^2\left(1 + \frac{c_3(c_3-1)}{3(2c_3-1)(2c_3-3)} \frac{w_0^2\Phi_{3,0}^2\ell^2}{N_c} v^{2c_3 - 3} + \OO(v, v^{2(2c_3 - 3)}) \right) &, \, \,\, c_3 > \frac{3}{2}
\end{array}
\right.
\end{equation}
\begin{equation}
\label{T0n7} h(u) =
\left\{
\begin{array}{ll}
h_{0}\left[1 + \frac{9-6c_3}{2(c_3-1)}v^{2c_3-2}\left(1 + \frac{w_0^2A^1_{z,1}\ell}{N_c} v^{3-2c_3} + \OO(v^{2c_3-2},v^{2(3-2c_3)})\right) \right]\! &,  1< c_3 < \frac{3}{2} \\
h_{0}\left[1 + \frac{c_3}{4(2c_3-1)} \frac{w_0^2\Phi_{3,0}^2\ell^2}{N_c} v^{2c_3 - 2}\left(1 + \OO(v, v^{2c_3 - 3})\right) \right]\! &,  c_3 > \frac{3}{2}
\end{array}
\right.
\end{equation}
\begin{equation}
\label{T0n8} \Phi(u) =
\left\{
\begin{array}{ll}
\frac{2\sqrt{N_c f_0}}{w_0 \ell}\ex^{A_0} v\left(1 + \frac{w_0^2A^1_{z,1}\ell}{2N_c(c_3-1)} v^{3-2c_3} + \OO(v^{2c_3-2},v^{2(3-2c_3)})\right) &, \, 1< c_3 < \frac{3}{2} \\
\frac{2\sqrt{N_c f_{0}}}{w_0\ell}\ex^{A_0} v\left(1 + \frac{c_3(c_3-1)}{6(2c_3-1)(2c_3-3)} \frac{w_0^2\Phi_{3,0}^2\ell^2}{N_c} v^{2c_3 - 3} + \OO(v, v^{2(2c_3 - 3)}) \right) &, \,  c_3 > \frac{3}{2}
\end{array}
\right.
\end{equation}
\begin{equation}
\label{T0n9} \Phi_3(u) =
\left\{
\begin{array}{ll}
\frac{\sqrt{6N_c f_0}}{w_0\ell}\ex^{A_0}v^{c_3}\left(1 +   \frac{w_0^2A^1_{z,1}\ell\, c_3}{2N_c(c_3-1)} v^{3 - 2c_3} + \OO(v^{2c_3-2}, v^{2(3-2c_3)}) \right) &\!\!\!\!\!\!\!\!, 1< c_3 < \frac{3}{2} \\
\sqrt{f_{0}}\Phi_{3,0}\ex^{A_0}v^{c_3}\left(1 + \frac{c_3^2(c_3-1)}{6(2c_3-1)(2c_3-3)} \frac{w_0^2\Phi_{3,0}\ell}{N_c} v^{2c_3 - 3} + \OO(v, v^{2(2c_3 - 3)}) \right)\! &,  c_3 > \frac{3}{2}
\end{array}
\right.
\end{equation}
\begin{equation}
\label{T0n10} A^1_z(u) =
\left\{
\begin{array}{ll}
\!\frac{2}{\ell}\ex^{A_0}\sqrt{3c_3(c_3-1)h_{0}}\bigg[1 - \frac{3-2c_3}{4c_3(c_3-1)^2}\frac{N_c}{w_0^2} v^{2c_3-2}\times &\!\!\!\!\!\!\!\!\!\!\!\!\!, 1< c_3 < \frac{3}{2}\\
\qquad\qquad\qquad \times \left(1 + \frac{w_0^2}{N_c}A^1_{z,1} v^{3-2c_3} + \OO(v^{2c_3-2},v^{2(3-2c_3)})\right) \bigg] \\
\!\frac{2}{\ell}\ex^{A_0}\sqrt{3c_3(c_3-1)h_{0}}\left(1 - \frac{\Phi_{3,0}^2\ell^2}{24(c_3-1)(2c_3-1)} v^{2c_3 - 2} + \OO(v^2, v^{2(2c_3 - 3)+1}) \right)\! &\!, c_3 > \frac{3}{2}
\end{array}
\right.
\end{equation}
where we defined the variable
\begin{equation}
\label{T0n11} v \equiv 2\ex^{A_{0}}\sqrt{\frac{3}{f_{0}}}\frac{r_H}{\ell} u \, .
\end{equation}
The expansions \eqref{T0n5}-\eqref{T0n10} can be systematically extended to higher orders.

\begin{figure}[h!]
	\begin{center}
		\includegraphics[scale=0.85]{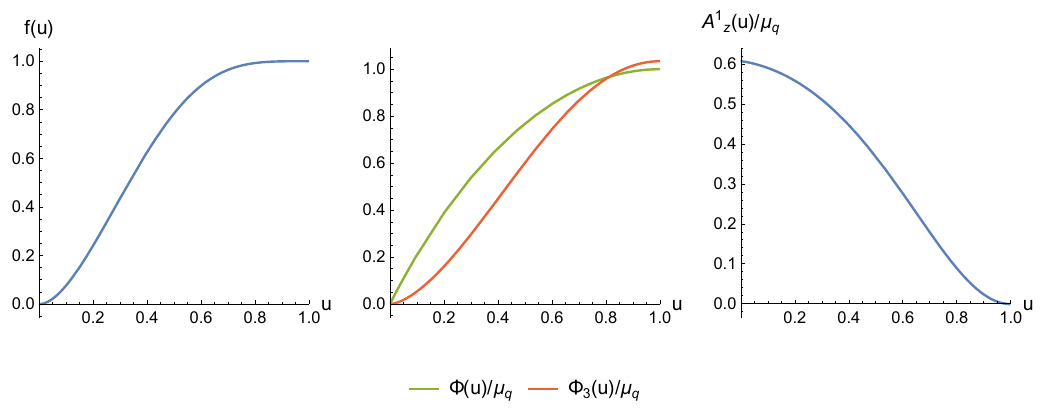}
		\caption{Profiles for the bulk fields in the extremal hairy black hole solution, which corresponds to the (0,1) phase at zero temperature, for $w_0 = 1$ and $n_3/n_q \simeq 2.05$. We show the blackening function on the left, the quark number (green) and isopin (red) gauge fields in the middle, and the non-abelian condensate on the right. The gauge fields are measured in units of the quark number chemical potential $\m_q$, and the coordinate $u$ is defined as $u = (r_H-r)/r_H$, such that the horizon lies at $u = 0$, and the AdS boundary at $u = 1$.}
		\label{fig:HBnq}
	\end{center}
	\begin{center}
		\includegraphics[scale=0.85]{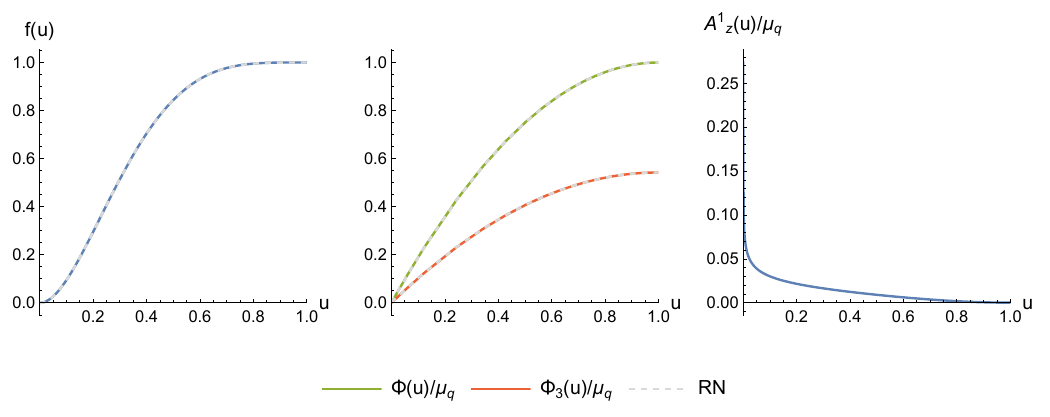}
		\caption{Same as figure \ref{fig:HBnq}, but for $n_3/n_q$ about 5\% above the transition point to the RN phase $(n_3/n_q)_c \simeq 0.56$. The gray dashed lines indicate the profiles for the fields in the RN solution with the same chemical potentials.}
		\label{fig:HBnqO2tr}
	\end{center}
\end{figure}

The existence of two branches of solutions with different IR asymptotics \eqref{T0n5}-\eqref{T0n10}, implies qualitative changes for some of the properties of the zero-temperature superconductor as $c_3 = 3/2$ is crossed. In particular, the derivative of the entropy density $\partial_T s$ at zero temperature is finite for $c_3 \geq 3/2$, whereas it becomes infinite at $c_3 < 3/2$ (see figure \ref{fig:sT} in appendix \ref{Thc32}). This naturally raises the question whether a phase transition happens at this point. The numerical results shown in appendix \ref{Thc32} (figure \ref{fig:m3}) however indicate that the density ratio $n_3/n_q$ remains smooth for all values of $c_3$, so that a finite order transition is unlikely. It remains plausible that an infinite order transition may happen (similar to the Berezinski-Kosterlitz-Thouless (BKT) transition), although determining whether it is indeed the case in this setup would require further investigation.

\subsubsubsection*{General properties of the solutions}

With the IR boundary conditions given as in \eqref{T0n5}-\eqref{T0n10}, the zero temperature condensed solution can be computed numerically for given values of $w_0$ and $n_3/n_q$. In figure \ref{fig:HBnq}, we show the resulting profiles for the blackening factor and the gauge fields, at $w_0 = 1$ and $n_3/n_q \simeq 2.05$. Those figures illustrate the main characteristics of the solution. First of all, it exhibits an extremal horizon, which is visible from the (double) zero of $f(u)$ at $u = 0$. This of course is something that we imposed from the beginning in the ansatz, since we knew that the baryon number has to be fractionalized in this model. This last point also implies that the baryon number flux at the horizon $\Phi'(0)$ should be non-zero, which is indeed observed in the middle figure \ref{fig:HBnq}, consistently with \eqref{T0n2}. Another consequence of equation  \eqref{T0n2}, is that the isospin flux $\Phi_3'(0)$ vanishes. This means that, unlike baryon number, the isospin charge is contained in cohesive states. Those states are precisely the quark pairs that constitute the p-wave condensate, dual to the non-abelian hair $A^1_z(u)$ shown in the right figure \ref{fig:HBnq}.

For generic values of $w_0$ and $n_3/n_q$, the general qualitative properties of the solution shown in figure \ref{fig:HBnq} remain the same. For each value of the ratio $n_3/n_q$, there is however a value of $w_0$ beyond which the solution becomes sub-leading to extremal RN (as expected from the phase diagram, figure \ref{fig:PDnq4d}), and eventually stops existing. When the transition is second order, the two happen at the same time, whereas there is a small difference between the dominance and existence bounds in the first order case.

It is interesting to observe how the condensed solution evolves when the (quantum) transition is approached. This behavior depends on the order of the transition, which changes along the transition line as shown in figure \ref{fig:PDnq4d_Tx} (furthest to the right). In particular, a first order transition is seen to occur only for $n_q$ much smaller than $n_3$, where the solution approaches the $n_q = 0$ solution that will be discussed later. For now we  focus on the case where $n_3/n_q$ is of order 1, and the transition is second order. In this regime, which corresponds to $w_0 \lesssim 2$, the qualitative properties of the solution are the same all along the transition line. For illustration, we  consider fixing $w_0 = 1$, and discuss the behavior of the solution when the critical value $(n_3/n_q)_c \simeq 0.56$ is approached from above.

\begin{figure}[h!]
	\begin{center}
		\includegraphics[scale=0.85]{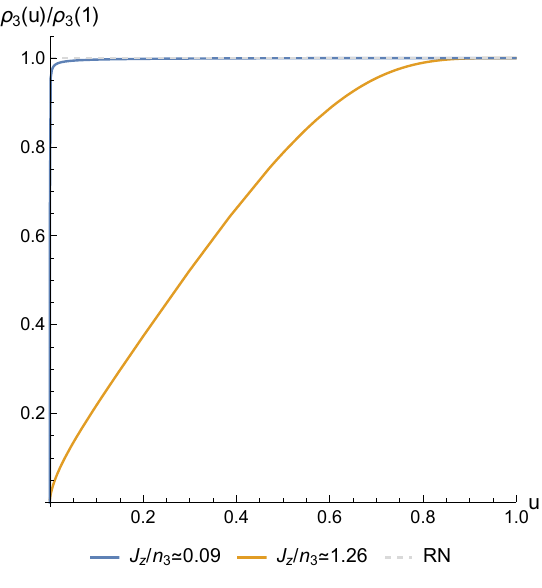}
		\caption{Ratio of the isospin charge contained within radius $u$, $\rho_3(u)$ \eqref{rho3def}, to the total bulk charge $\rho_3(1)$, for the solution shown in figure \ref{fig:HBnqO2tr} (blue), and figure \ref{fig:HBnq} (orange). The gray dashed line shows the behavior in the RN solution, where the absence of the isospin-charged hair implies that $\rho_3(u)$ is a constant equal to the black hole charge. }
		\label{fig:rho3}
	\end{center}
\end{figure}

Figure \ref{fig:HBnqO2tr} shows the fields of the superconducting solution for $w_0 = 1$ and $n_3/n_q$ about 5\% above the critical point. The gauge fields and blackening function are observed to be very close to the RN solution, whereas the non-abelian hair is approaching zero. Note however that, from \eqref{T0n10}, $(A^1_z)'(r_H)$ is infinite whenever $c_3<3/2$, which includes the solution of figure \ref{fig:HBnqO2tr}. This behavior implies that, as the transition is approached, the bulk isospin density tends towards a Dirac delta at the horizon, so that the bulk isospin charge is contained within an increasingly thin shell just above the horizon. In order to analyze this, we define the bulk density through
\begin{equation} \label{rho3def}
 \rho_3(r) = \frac{e^{A(r)}\sqrt{h(r)}\Phi_3'(r)}{2\ell} \, ,
\end{equation}
where the normalization was chosen such that $\rho_3(r=0)$ at the boundary equals the coefficient $\Phi^{(1)}$ in~\eqref{d31}. For the solution shown in figure \ref{fig:HBnqO2tr}, figure \ref{fig:rho3} shows that more than 90\% of the isospin charge is already contained within $u\lesssim 0.02$.

\subsubsubsection*{Effective IR theory}

We now discuss the effective theory that emerges at low energies in the superconducting phase. As mentioned above, this IR theory is controlled by the near-horizon geometry of the condensed solution. The IR geometry of the extremal hairy black hole is the same as for extremal RN, that is AdS$_2 \times \mathbb{R}^3$ (see appendix \ref{appH} for the definition of the near-horizon limit). This means that the effective theory at low energy is described by a CFT$_1$, which is invariant under conformal transformations of the time coordinate. This CFT is similar to the effective IR theory that arises in the RN state. However, it differs by the effective sources that emerge for the IR operators at the AdS$_2$ boundary. Whereas the RN CFT contains a source for the isospin gauge field $\Phi_3$, which corresponds to the effective isospin chemical potential, that source vanishes in the paired phase, where it is instead the condensing field $A^1_z$ which is sourced.

This difference in the sources has implications for the spectrum of the IR CFT, and in particular for the operators dual to the chiral gauge fields. As explained in appendix \ref{appH}, in the RN phase, most gauge fields are massless in AdS$_2$, with IR conformal dimension $\D = 1$. The only exception is for the condensing gauge field $A^1_z$, whose conformal dimension decreases with isospin density
\begin{equation}
\label{T0n12} \D^1_z = \frac{1}{2} + \frac{1}{2}\sqrt{1 - \frac{4N_c}{3w_0^2}\frac{(n_3/n_q)^2}{1 + (n_3/n_q)^2}} \, .
\end{equation}
This means that the dual operator becomes more and more relevant as $n_3/n_q$ is increased. For low enough $w_0$, equation \eqref{T0n12} implies that a phase transition should happen when $n_3/n_q$ is increased beyond some critical value, since unitarity will be violated in the IR CFT as soon as the term below the square root becomes negative. The critical line where this happens obeys
\begin{equation}
\label{Tn012b} \frac{n_3}{n_q} = \left(\frac{4N_c}{3w_0^2} -1\right)^{-\frac{1}{2}} \, ,
\end{equation}
and corresponds to the saturation of the AdS$_2$ Breitenlohner-Freedman (BF) bound $m^2\ell_2^2 = -1/4$, by the mass of the $A^1_z$ mode. For $n_3/n_q$ larger than \eqref{Tn012b}, the AdS$_2$ IR geometry develops an instability, which should result in a phase transition.

\begin{figure}[h!]
	\begin{center}
		\includegraphics[scale=0.85]{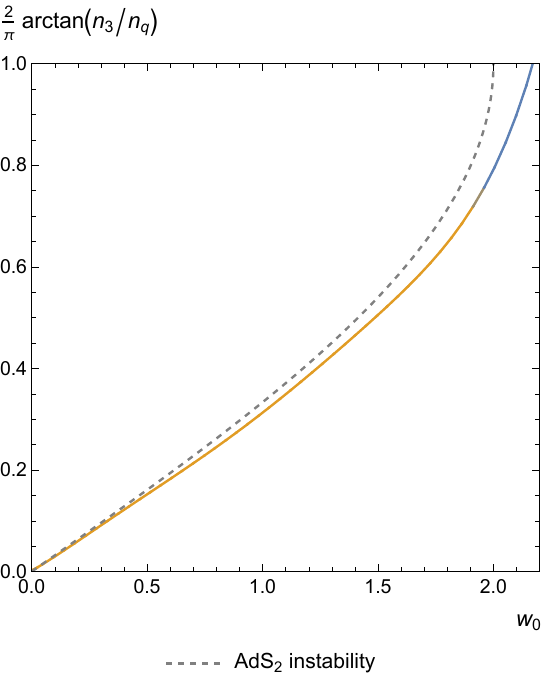}
		\caption{Phase transition line at zero temperature. Orange corresponds to a second order transition, and blue to first order. For comparison, the gray dashed line shows the onset of the AdS$_2$ BF instability \eqref{Tn012b}.}
		\label{fig:plT0}
	\end{center}
\end{figure}

The onset of the AdS$_2$ instability \eqref{Tn012b} gives an upper bound on the critical value of $n_3/n_q$ where the transition to the (0,1) phase takes place, but the two need not coincide. In figure \ref{fig:plT0}, the numerical zero-temperature phase transition line is compared with \eqref{Tn012b}. This shows that the transition line agrees with \eqref{Tn012b} only in the limit of zero isospin density, where the condensed phase disappears all together. Therefore, the actual instability which triggers the formation of the condensate is not normalizable in AdS$_2$. This is consistent with the properties of the condensed solution mentioned above : the AdS$_2$ IR geometry is preserved, but the sources at the AdS$_2$ boundary are modified compared with the RN case.

The spectrum of the IR CFT is strongly affected by the phase transition. An analysis completely analogous to appendix \ref{appH} shows that, in the condensed phase, the situation is reversed for the isospin and condensate gauge fields compared with RN. That is, the perturbations of the hair $A^1_z$ become massless, whereas the isospin gauge field $\Phi_3$ gets a non-zero mass. Because this mass is real, the isospin charge is irrelevant in the condensed phase, with a conformal dimension
\begin{equation}
\label{T0n13} \D_3 = c_3 \, .
\end{equation}

\begin{figure}[h!]
	\begin{center}
		\includegraphics[scale=0.85]{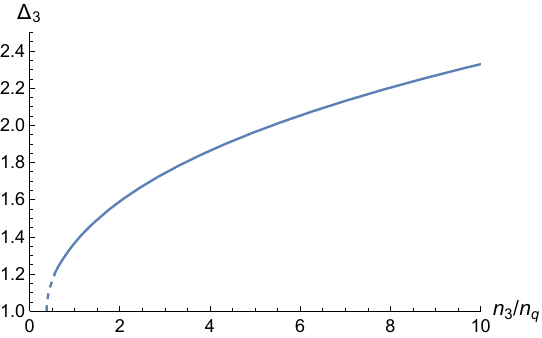}
		\caption{IR scaling dimension of the isospin charge as a function of the density ratio $n_3/n_q$, in the condensed phase and at $w_0 = 1$. The full line shows the points that were computed numerically. Those results reveal a square-root behavior close to the transition to RN, which was used to extrapolate down to the RN point $\D_3=1$ (dashed line).}
		\label{fig:D3IR}
	\end{center}
\end{figure}

For a given $w_0$, computing the solution numerically makes it possible to determine the relation between $\D_3$ and the phase parameter $n_3/n_q$. The result is shown in figure \ref{fig:D3IR} for $w_0 = 1$, where it is observed that the isospin charge becomes more and more irrelevant as we go further into the condensed phase, towards zero quark density. Physically, this is due to the isospin charge being screened by the condensate.

\subsubsection{Zero quark number density}

We  now consider the limiting case where the quark density $n_q$ is taken to vanish in the zero temperature hairy black hole solution. Since the extremal horizon is charged only under quark number - the isospin charge being carried by the hair - it is expected that the solution becomes horizonless at $n_q = 0$. This should correspond to a solitonic configuration similar to the solution analyzed in \cite{Horowitz09}. Moreover, figure \ref{fig:D3IR} indicates that the IR scaling of the isospin gauge field goes to infinity as $n_q$ goes to zero, which suggests an exponential IR scaling of the fields in this limit
\begin{equation}
\label{T01} \f(r) \underset{r\to\infty}{=} \sum_{j \in \mathbb{N}} \f_j(r) \ex^{-j\, d_\f \frac{r}{\ell}} \sp \f\in\{A,f,h,\Phi,\Phi_3,A^1_z\} \, ,
\end{equation}
where the exponents $d_\f$ are a priori unknown, and the coefficients $\f_j(r)$ have power-law asymptotics as $r$ goes to infinity.

Substituting this ansatz in the near-horizon equations of motion reveals that there is indeed a solution with these asymptotics. At leading order the IR geometry is AdS$_5$
\begin{equation}	
\label{T02} A(r) = -\log{\left(\frac{r-r_0}{\ell \sqrt{f_0}}\right)} + \dots \sp f(r) = f_0  + \dots \sp h(r) = h_0 + \dots \, ,
\end{equation}
up to exponentially suppressed corrections at large $r$ that are indicated by the dots. The parameter $r_0$ reflects the invariance of the equations of motion under shifts of the holographic coordinate. 
At $r\to \infty$, the gauge fields then behave as fields in AdS$_5$, for which non-zero sources are turned on. Specifically, $A^1_z$ behaves as a massless gauge field, for which the regular solution is a constant
\begin{equation}
\label{T03} A^1_z(r) = A^1_{z,0} + \dots \, ,
\end{equation}
whereas $\Phi_3$ obeys the equation for a massive gauge field in AdS$_5$, whose mass is controlled by $A^1_{z,0}$
\begin{equation}
\label{T04} \Phi_3''(r) - \frac{1}{r-r_0}\Phi_3'(r) - \frac{(A^1_{z,0})^2}{f_0 h_0}\Phi_3(r) + \dots = 0 \, .
\end{equation}
Equation \eqref{T04} has a well-known regular solution in terms of a modified Bessel function
\begin{equation}
\label{T05} \Phi_3(r) = \Phi_{3,0} \a_3\bar{r} K_1\left(\a_3\bar{r}\right) + \dots \sp \bar{r} \equiv \frac{r-r_0}{\ell \sqrt{f_0}} \sp \a_3 \equiv \frac{A^1_{z,0}\ell}{\sqrt{h_0}} \, .
\end{equation}
For an asymptotically large argument $x$, the Bessel function $K_1(x)$ decays as \\
$\sqrt{\pi/(2x)} \ex^{-x}$ times an analytic series in $x^{-1}$. Hence, \eqref{T05} gives the exact expression for the coefficient $\Phi_{3,1}(r)$ at leading exponential order in the series expansion at $r \to\infty$, \eqref{T01}.

From the result for $\Phi_3$ \eqref{T05}, the first exponential corrections $\f_1(r)$ to the other fields $A,f,h$ and $A^1_z$ can be written as integrals of Bessel functions. This writing is not very illuminating though. We shall instead provide the leading terms in the series expansion of $\f_1(r)$ in powers of $r^{-1}$. The latter can be obtained by substituting the $r\to \infty$ limit of $\Phi_3(r)$ into the IR equations of motion, which gives the following result
\begin{equation}
\label{T06} A(r) = -\log{\bar{r}} - \frac{\pi w_0^2}{48N_c}\frac{\Phi_{3,0}^2\ell^2}{f_0}\a_3\bar{r}^3 \ex^{-2\a_3\bar{r}} \left(1 + \OO(1/\bar{r})\right) + \OO(\ex^{-4\a_3\bar{r}}) \, ,
\end{equation}
\begin{equation}
\label{T07} f(r) = f_0\left(1 + \frac{\pi w_0^2}{8N_c}\frac{\Phi_{3,0}^2\ell^2}{f_0}\a_3\bar{r}^3 \ex^{-2\a_3\bar{r}} \left(1 + \OO(1/\bar{r})\right) + \OO(\ex^{-4\a_3\bar{r}}) \right) \, ,
\end{equation}
\begin{equation}
\label{T08} h(r) = h_0\left(1 + \frac{\pi w_0^2}{16N_c}\frac{\Phi_{3,0}^2\ell^2}{f_0}\a_3\bar{r}^3 \ex^{-2\a_3\bar{r}} \left(1 + \OO(1/\bar{r})\right) + \OO(\ex^{-4\a_3\bar{r}}) \right) \, ,
\end{equation}
\begin{equation}
\label{T09} A^1_z(r) = \frac{\sqrt{h_0}}{\ell}\left(\a_3 - \frac{\pi}{8}\frac{\Phi_{3,0}^2\ell^2}{f_0}\bar{r}\ex^{-2\a_3\bar{r}} \left(1 + \OO(1/\bar{r})\right) + \OO(\ex^{-4\a_3\bar{r}}) \right) \, ,
\end{equation}
where the variable $\bar{r}$ was defined in \eqref{T05}. The IR expansions \eqref{T06}-\eqref{T09} contain four\footnote{The parameter $r_0$ in \eqref{T02} is in principle an additional parameter when working in conformal coordinates, which is fixed by defining the boundary to lie at $r=0$.} dimensionless parameters $f_0,h_0,\a_3$ and $\Phi_3\ell^2$. Three of these correspond to rescalings, and the last degree of freedom is fixed by requiring the source to vanish for $A^1_z$ at the (UV) AdS boundary. Thus, we are left with no parameters, consistently with the regime that we are considering, at $T=n_q=0$.

\begin{figure}[h!]
	\begin{center}
		\includegraphics[scale=0.85]{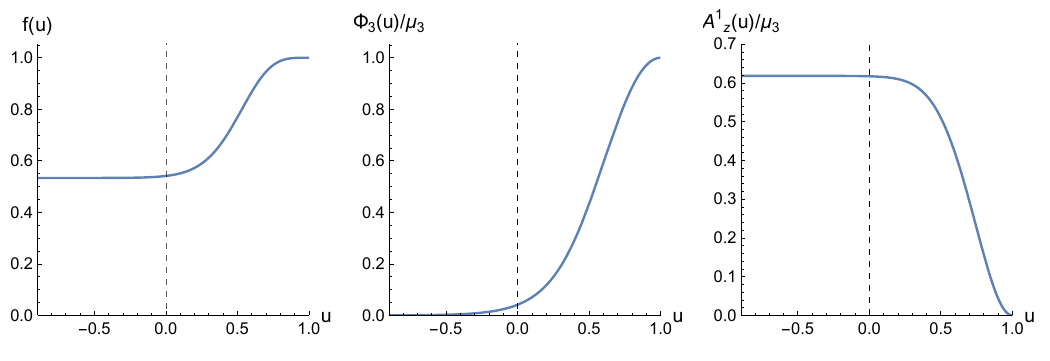}
		\caption{Profiles for the bulk fields in the soliton solution, which corresponds to the (0,1) phase at zero temperature and quark number density, for $w_0 = 1$. We show the blackening function on the left, the isospin gauge field in the middle, and the non-abelian condensate on the right. The gauge fields are measured in units of the isospin chemical potential $\m_3$, and the coordinate $u$ is defined as $u = (r_H-r)/r_H$, with $r_H = 2\sqrt{3N_c}/(w_0\m_3)$, the radius of the extremal RN horizon with the same isospin chemical potential. The boundary is at $u=1$, and the dashed line at $u=0$ indicates where the RN horizon would lie. The soliton solution does not have a horizon there though, and the solution extends to $u \to -\infty$.}
		\label{fig:Sw01}
	\end{center}
	\begin{center}
		\includegraphics[scale=0.85]{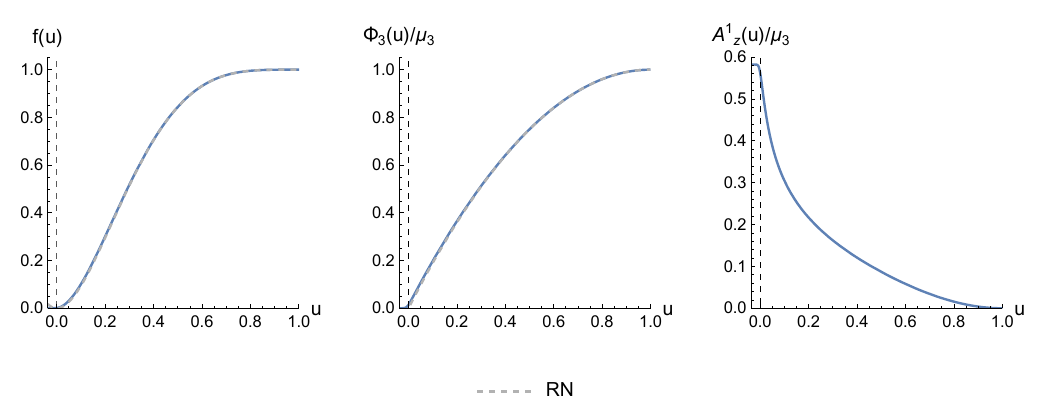}
		\caption{Same as figure \ref{fig:Sw01}, but for $w_0\simeq 2.17$, where the free energies of the (0,1) and RN phases coincide. The gray dashed lines indicate the profiles for the fields in the RN solution with the same chemical potential $\m_3$.}
		\label{fig:SO1}
	\end{center}
\end{figure}

The solution for the superconducting state at zero temperature and quark number density can be computed by solving numerically the equations of motion \eqref{B3}-\eqref{B8}, with the boundary conditions given in the IR by \eqref{T05}-\eqref{T09}. The resulting profiles for the blackening function and the gauge fields  at $w_0 = 1$ are shown in figure \ref{fig:Sw01}. The solution is seen to take the form of a domain wall, whose size and location are set by the only available scale, that is the isospin density. From the field theory point of view, this solution corresponds to an RG flow which interpolates between a UV and an IR fixed point, that are both described by a four-dimensional CFT. As was the case at finite $n_q$, the screening of the isospin charge by the condensate translates into the isospin gauge field going from massless in the UV to massive in the IR. The condensate gauge field $A^1_z$, meanwhile, becomes less relevant along the flow, its negative mass term going to zero in the IR.

\begin{figure}[h!]
	\begin{center}
		\includegraphics[scale=0.85]{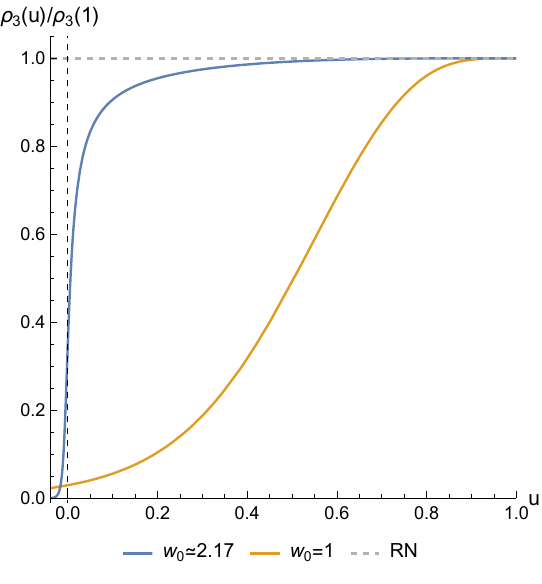}
		\caption{Ratio of the isospin charge contained within radius $u$, $\rho_3(u)$, to the total bulk charge $\rho_3(1)$, for the solution shown in figure \ref{fig:SO1} (blue), and figure \ref{fig:Sw01} (orange). The definition of the coordinate $u$ is given in the caption of figure \ref{fig:SO1}, and depends on $w_0$. The gray dashed line shows the behavior in the RN solution, where $\rho_3(u)$ is a constant equal to the black hole charge. The black dashed line at $u = 0$ indicates the location of the horizon in the RN solution with the same isospin chemical potential $\m_3$.}
		\label{fig:rho3O1}
	\end{center}
\end{figure}

The approach to the phase transition with the uncondensed RN phase is qualitatively different from the $n_q \neq 0$ solutions of figures~\ref{fig:HBnq} and~\ref{fig:HBnqO2tr}, since the transition is now first order. To study this in more detail, we show the zero density solution at $w_0 \simeq 2.17$ in figure \ref{fig:SO1}, which is the point where the two phases have equal free energy. The geometry and isospin gauge field are observed to essentially match the RN solution, with in particular the emergence of an extremal horizon at the same location as RN. However, the condensate gauge field is clearly non-zero, and extends away from the horizon. As shown in figure \ref{fig:rho3O1}, this implies that a significant fraction of the isospin charge is sitting outside of the RN horizon. Hence, the superconducting solution at the transition approaches an extremal hairy black hole, whose charge is split between the horizon and the hair. As $w_0$ is increased beyond the transition point, the flavor Yang-Mills coupling $g \propto w_0^{-1}$ keeps getting smaller and smaller compared with the gravitational attraction. Above a certain value $w_0^*$, there is no stable solution able to support a hair anymore, and the soliton collapses to the RN black hole. Numerically, $w_0^*$ is found to lie very close to the first order transition point $w_{0,c} \simeq 2.17$, such that $(w_0^*-w_{0,c})/w_{0,c}\simeq 0.005$. The phase transition may happen at $w_0^*$, or somewhere between $w_{0,c}$ and $w_0^*$, depending on dynamics. Whenever it happens, it will be realized by the gravitational collapse of the solution.

\section*{Acknowledgements}\label{ACKNOWL}
\addcontentsline{toc}{section}{Acknowledgements}

We would like to thank Javier Subils for discussions.

This work is partially supported by the European MSCA grant HORIZON-MSCA-2022-PF-01-01 ``BlackHoleChaos" No.101105116,  by the In2p3 grant ``Extreme Dynamics", by the H.F.R.I call ``Basic research Financing (Horizontal support of all Sciences)" under the National Recovery and Resilience Plan ``Greece 2.0" funded by the European Union -- NextGenerationEU (H.F.R.I. Project Number: 15384.) and by a Humboldt-Siemens research award. MJ has been supported by an appointment to the JRG Program at the APCTP through the Science and Technology Promotion Fund and Lottery Fund of the Korean Government. He has also been supported by the Korean Local Governments -- Gyeong\-sang\-buk-do Province and Pohang City -- and by the National Research Foundation of Korea (NRF) funded by the Korean government (MSIT) (grant number 2021R1A2C1010834).

\newpage

\appendix
\renewcommand{\theequation}{\thesection.\arabic{equation}}
\addcontentsline{toc}{section}{Appendix\label{app}}
\section*{Appendix}

\section{The AdS Reissner-Nordstr\"om solution}

\label{Sec:bgd}

We review in this appendix the derivation of the Reissner-Nordström black hole solution in AdS$_{d+1}$, with $d \geq 3$.

The equations of motion from the action \eqref{Sb} are the Einstein-Yang-Mills equations
\begin{align}
\nn R_{MN} - \frac{1}{2}\left(R + \frac{d(d-1)}{\ell^2}\right)g_{MN} = -\frac{w_0^2\ell^2}{4N_c} \text{Tr}\bigg\{\mathbf{F}^{(L)}_{\hphantom{(L)}MP} &\mathbf{F}^{(L)P}_{\hphantom{(L)P}N} + \frac{1}{4}\mathbf{F}^{(L)}_{PQ}\mathbf{F}^{(L)PQ}g_{MN}  +\\
\label{Ba1} & + (L \leftrightarrow R) \bigg\} \, ,
\end{align}
\be
\label{Ba2} D^{(L/R)}_M\Big(\sqrt{-g}F^{(L/R)MN}\Big) = 0 \, ,
\ee
with $D^{(L/R)}_M$ the Yang-Mills covariant derivatives
\be
\label{Ba3} D^{(L)}_M \equiv \partial_M - i [L_M,\, . \,\,] \sp D^{(R)}_M \equiv \partial_M - i [R_M,\, . \,\,] \, .
\ee
The background solution is found by starting from the ansatz
\be
\label{Ba4} \mathrm{ds}^2 = \ex^{2A(r)}\left( -f(r)\mathrm{dt}^2 + f(r)^{-1}\mathrm{dr}^2 + \vec{\mathrm{d x}}^2 \right) \, ,
\ee
\be
\label{Ba5} \mathbf{R}_M = \mathbf{L}_M = \frac{1}{2}\d^{0}_M\big(
\Phi(r)\mathbb{I}_2 +  \Phi_3(r) \s_3  
\big) \, .
\ee
This ansatz fixes the gauge for the gauge field, up to a shift by a constant. As we shall see below, the regular boundary conditions in the IR \eqref{Ba9} remove this degeneracy.

Substituting the ansatz \eqref{Ba4}-\eqref{Ba5} into the equations of motion \eqref{Ba1}-\eqref{Ba2} results in the following system of equations for the ansatz fields
\be
\label{Ba6} \partial^2_r A - (\partial_r A)^2 = 0 \, ,
\ee
\be
\label{Ba6b} \partial_r A\big(\partial_r f + d\partial_r A f(r)\big) - \frac{d}{\ell^2}\ex^{2A(r)}  
+ \frac{w_0^2\ell^2}{12N_c}\ex^{-2A(r)}\bigg((\partial_r \Phi_3)^2 + (\partial_r \Phi)^2\bigg) = 0 \, ,
\ee
\be
\label{Ba7} 
\partial_r\Big(\ex^{(d-3)A(r)} \partial_r \Phi\Big) = 0 \sp \partial_r\Big(\ex^{(d-3)A(r)} \partial_r \Phi_3\Big) = 0  \, .
\ee
The two integration constants of \eqref{Ba6} correspond to translations and rescalings of $r$. We fix the definition of the coordinate $r$ by writing the solution as
\be
\label{Ba8} A(r) = \log{\left(\frac{\ell}{r}\right)} \, ,
\ee
which implies in particular that the boundary is located at $r=0$. We look for a solution with a horizon at $r=r_H$, where the blackening function $f(r)$ vanishes. For the gauge field to be regular at the horizon, the time component should vanish
\be
\label{Ba9} \Phi(r_H) = \Phi_3(r_H) = 0 \, .
\ee
This implies that the solutions of \eqref{Ba7} are given by
\be
\label{Ba10} \Phi = \mu_q\left(1 - \left(\frac{r}{r_H}\right)^{d-2}\right) \sp \Phi_3 = \mu_3\left(1 - \left(\frac{r}{r_H}\right)^{d-2}\right) \, ,
\ee
with the boundary sources $\m_q$ and $\m_3$ corresponding respectively to the quark number and isospin chemical potentials. Finally, the solutions for the gauge fields and the scale factor $A(r)$ can be substituted in \eqref{Ba6b}, which yields an equation for $f(r)$
\be
\label{Ba11} \partial_r f -\frac{d}{r}(f(r)-1) - \frac{(d-2)^2w_0^2}{4(d-1)N_c}\mu^2 r \left(\frac{r}{r_H}\right)^{2d-4} = 0 \, ,
\ee
where $\m\equiv \sqrt{\m_q^2 + \m_3^2}$. The solution takes the form
\be
\label{Ba12} f(r) = 1 - \left(\frac{r}{r_H}\right)^d\left(1 + \frac{(d-2)w_0^2}{4(d-1)N_c}\m^2 r_H^2\right) + \frac{(d-2)w_0^2}{4(d-1)N_c}\m^2 r_H^2\left(\frac{r}{r_H}\right)^{2d-2} \, ,
\ee
where we fixed the constant of integration such that $f$ approach one at the boundary.
To avoid a conical singularity of the Euclidean solution at finite temperature, the derivative of $f(r)$ at the horizon should be related to the field theory temperature
\be
\label{Ba13} f'(r_H) = -4\pi T \, .
\ee
This condition results in an equation for the horizon radius $r_H$
\be
\label{Ba14} \frac{(d-2)^2w_0^2}{4(d-1)N_c} \m^2r_H^2 = d - 4\pi T r_H \, ,
\ee
whose solution determines the location of the black-hole horizon as a function of the chemical potential $\m$ and the temperature
\be
\label{Ba15} r_H(T,\m) = \frac{d}{2\pi T} \left( 1 + \sqrt{1 + \frac{d(d-2)^2w_0^2}{16(d-1)N_c}\frac{\mu^2}{\pi^2T^2}} \right)^{-1} \, .
\ee
Note that \eqref{Ba14} allows to rewrite $f(r)$ in a more compact form
\begin{equation}
\label{Ba16} f(r) = 1 - \left(\frac{r}{r_H}\right)^d\left(1 + \frac{d-4\pi Tr_H}{d-2}\right) +\frac{d-4\pi Tr_H}{d-2}\left(\frac{r}{r_H}\right)^{2d-2}  \, .
\end{equation}

\section{The Yang-Mills constraints near the boundary}

\label{AppA}

In this appendix, we detail the near-boundary analysis of the constraints that the Yang-Mills equations impose on the gauge fields. We consider a general number of boundary dimensions $d\geq 3$.

Near the boundary, the bulk geometry asymptotes that of AdS, and the metric can be written as
\begin{equation}
\label{A1} \intd s^2 = \frac{\ell^2}{r^2}\big(\intd r^2 + g_{\m\n}(r^2)\intd x^\m\intd x^\n\big) \, ,
\end{equation}
which is the so-called Fefferman-Graham gauge \cite{FG}. We consider the case where $g_{\m\n}$ is diagonal, and becomes flat at the boundary $g_{\m\n}(0) = \eta_{\m\n}$, so that the near-boundary behavior is of the form \cite{deHaro00}
\begin{equation}
\label{A2} g_{\m\n}(r) = \eta_{\m\n} + t_{\m\n}r^d\big(1 + \OO(r^{d-2}) \big) \, ,
\end{equation}
where $t_{\mu\nu}$ is proportional to the expectation value of the stress tensor.

For the gauge fields, the situation considered in this work is such that only two constant sources are turned on a the boundary, $\mu_q$ and $\mu_3$, which correspond to the quark number and isospin chemical potentials. As discussed in section \ref{OP}, the symmetries of the theory make it possible to restrict the condensing gauge field to the components $A^1_x,A^2_y,A^1_t,A^2_t$. Here, we show that the Gauss law constraint in the radial gauge $A_r = 0$, imposes that the time components in the $1,2$ directions should also be zero.

In accordance with the discussion above, the expansion of the gauge fields close to the boundary is given by
\begin{equation}
\nn \Phi(r) = \m_q + \r_q r^{d-2} \big(1 + \OO(r^d)\big) \sp \Phi_3(r) = \m_3 + \r_3 r^{d-2} \big(1+ \OO(r^d)\big) \, ,
\end{equation}
\begin{equation}
\label{A3} A_x^1(r) = \r^1_x r^{d-2}\big(1 + \OO(r^2)\big) \sp A_y^2(r) = \r^2_y r^{d-2} (1 + \OO(r^2)) \, ,
\end{equation}
\begin{equation}
\nn A_t^{1,2}(r) = \r^{1,2}_t r^{d-2}\big(1 + \OO(r^2)\big) \, .
\end{equation}

Let us note at this point that \eqref{A3} is not the most general behavior for the gauge fields near the boundary. Indeed, in presence of sources (here the chemical potentials), the expansion generically contains all even powers of $r$, and also a term of order $\OO(r^{d-2}\log{r})$ when $d$ is even. However, those terms are zero for a vanishing field strength at the boundary, which explains the simplicity of the expansions in \eqref{A3}.

Now consider the YM equations of motion \eqref{PD2} for the non-abelian gauge fields
\begin{equation}
\label{A4} {1\over \sqrt{-g}}\pa_{M}(\sqrt{-g}F^{a, MN})+\e^{abc}A^a_MF^{b,MN} \, .
\end{equation}
Setting $A_r^a=0$ and $\pa_{\nu}A^a_{M}=0$, the equation in the direction $N=r$ yields a constraint for the gauge fields
\begin{equation}
\label{A5} \e^{abc}A_{\m}^{b}\pa_r A^{c,\mu}=0 \, .
\end{equation}
From \eqref{A3}, the leading piece near the boundary of equation \eqref{A5} is ${\cal O}(r^{d-3})$ and implies
\begin{equation}
\label{A6} \mu_3 \r^{1,2}_{t}=0 \, .
\end{equation}
Therefore, as long as $\mu_3\neq 0$, the time components $A^{1,2}_t(r)$ have to be identically zero.

\section{Equations of motion}

\label{AppB}

We present in this appendix the equations of motion obeyed by the ansatz fields for the two condensed solutions presented in the text. Those equations are obtained by substituting the appropriate ansatz derived in section~\ref{OP} in the Einstein-Yang-Mills equations \eqref{PD1}-\eqref{PD2}. Since we studied both $d=2+1$ and $d=3+1$ boundary dimensions, we write the equations for general $d$.

We start from the solution labeled $(0,1)$, where a single component of the gauge field condenses. The solution exists for $d\geq 2$ boundary dimensions, and the appropriate ansatz is given by
\begin{equation}
\label{B1} \mathbf{L} = \mathbf{R} = \frac{1}{2} \Phi(r) \intd t\,\mathbb{I}_2 + \frac{1}{2} \Phi_3(r) \intd t\,\s^3+ \frac{1}{2} A^1_z(r) \intd z\, \s^1 \, ,
\end{equation}
\begin{equation}
\label{B2} \mathrm{ds}^2 = \ex^{2A(r)}\left( -f(r)\mathrm{dt}^2 + f(r)^{-1}\mathrm{dr}^2 + \intd x_1^2 + \dots + \intd x_{d-2}^2 + h(r) \intd z^2 \right) \, .
\end{equation}
The equations of motion obeyed by the ansatz fields are given by
\begin{equation}
\label{B3} A''(r) - A'(r)\left(A'(r)+\frac{h'(r)}{2h(r)}\right) - \frac{f'(r)h'(r)}{2(d-1)f(r)h(r)} + \frac{w_0^2\ell^2\ex^{-2A(r)}}{2(d-1)N_c} \frac{A^1_z(r)^2\Phi_3(r)^2}{f(r)^2h(r)} = 0 \, ,
\end{equation}
\begin{align}
\label{B4} h''(r)&+ h'(r) \left((d-1)A'(r)+\frac{f'(r)}{f(r)}-\frac{h'(r)}{2 h(r)}\right)+ \\
\nn &+\frac{w_0^2 \ell^2}{2 N_c}\ex^{-2 A(r)}\left( (A^1_z)'(r)^2 - \frac{A^1_z(r)^2 \Phi_3(r)^2}{f(r)^2}\right) = 0 \, ,
\end{align}
\begin{align}
\label{B5} &f(r) \left(\frac{A'(r) h'(r)}{h(r)}+d A'(r)^2 \right)+f'(r) \left(A'(r)+\frac{h'(r)}{2(d-1) h(r)}\right) -\frac{d\, \ex^{2 A(r)}}{\ell^2}- \\
\nn &-\frac{w_0^2 \ell^2}{4(d-1) N_c} \ex^{-2 A(r)} \left(\frac{f(r)}{h(r)}(A^1_z)'(r)^2 + \frac{A^1_z(r)^2 \Phi_3(r)^2}{f(r) h(r)}- \left(\Phi_3'(r)^2+\Phi'(r)^2\right)\right) = 0 \, ,
\end{align}
\begin{equation}
\label{B6} \Phi'(r) + \ex^{-(d-3)A(r)}\ell^{-2} \frac{\bar{n} }{ \sqrt{h(r)} } = 0 \, ,
\end{equation}
\begin{equation}
\label{B7} \Phi_3''(r)+ \Phi_3'(r) \left((d-3)A'(r)+\frac{h'(r)}{2 h(r)}\right)-\frac{A^1_z(r)^2 \Phi_3(r)}{f(r) h(r)} = 0 \, ,
\end{equation}
\begin{equation}
\label{B8} (A^1_z)''(r)+ (A^1_z)'(r) \left((d-3)A'(r)+\frac{f'(r)}{f(r)}-\frac{h'(r)}{2 h(r)}\right)+\frac{A^1_z(r) \Phi_3(r)^2}{f(r)^2} = 0 \, ,
\end{equation}
with $\bar{n}$ an integration constant.

We now discuss the $(1,1)$ solution, which exists for $d\geq 3$, and is such that two components of the gauge field condense with the same amplitude. The ansatz in this case is
\begin{equation}
\label{B9} \mathbf{L} = \mathbf{R} = \frac{1}{2} \Phi(r) \intd t\,\mathbb{I}_2 + \frac{1}{2} \Phi_3(r) \intd t\,\s^3+ \frac{1}{2} w(r) \big(\intd x\, \s^1 + \intd y\, \s^2\big) \, ,
\end{equation}
\begin{equation}
\label{B10} \mathrm{ds}^2 = \ex^{2A(r)}\left( -f(r)\mathrm{dt}^2 + f(r)^{-1}\mathrm{dr}^2 + \intd x^2 + \intd y^2 + h(r) (\intd z_1^2 + \dots + \intd z_{d-3}^2) \right) \, ,
\end{equation}
where the last term in the metric ansatz is present for $d\geq 4$. The equations of motion are given by
\begin{align}
\nn &A''(r) - A'(r)\left(A'(r)+(d-3)\frac{h'(r)}{2h(r)}\right) - \frac{d-3}{2(d-1)}\frac{h'(r)}{h(r)}\left(\frac{f'(r)}{f(r)} + (d-4)\frac{h'(r)}{2h(r)}\right) +\\
\label{B11} &\,\, + \frac{w_0^2\ell^2}{4N_c}\ex^{-2A(r)} \left(w'(r)^2 + \frac{w(r)^2}{(d-1)f(r)^2}\big((d-3)f(r)w(r)^2 - (d-5) \Phi_3(r)^2\big)\right) = 0 \, ,
\end{align}
\begin{align}
\nn &f'(r) \left(A'(r)+\frac{d-3}{2(d-1)}\frac{h'(r)}{h(r)}\right) -\frac{d\, \ex^{2 A(r)}}{\ell^2}+ \\
\nn & + f(r) \left[(d-3)\frac{h'(r)}{h(r)}\left(A'(r)  + \frac{d-4}{4(d-1)} \frac{h'(r)}{h(r)}\right) +d A'(r)^2 \right] -\\
\label{B13} &-\frac{w_0^2 \ell^2\ex^{-2 A(r)}}{4(d-1) N_c}  \left(2f(r)w'(r)^2 + 2\frac{w(r)^2 \Phi_3(r)^2}{f(r)}- \left(\Phi_3'(r)^2+\Phi'(r)^2 + w(r)^4\right)\right) = 0 \, ,
\end{align}
\begin{equation}
\label{B14} \Phi'(r) +  \big(\ex^{A(r)}\sqrt{h(r)}\big)^{-(d-3)}\ell^{-2} \bar{n} = 0 \, ,
\end{equation}
\begin{equation}
\label{B15} \Phi_3''(r)+ (d-3)\Phi_3'(r) \left(A'(r)+\frac{h'(r)}{2 h(r)}\right)-\frac{2w(r)^2 \Phi_3(r)}{f(r)} = 0 \, ,
\end{equation}
\begin{equation}
\label{B16} w''(r)+ w'(r) \left(\frac{f'(r)}{f(r)}+(d-3)\left(A'(r)+\frac{h'(r)}{2 h(r)}\right)\right)+\frac{w(r) \Phi_3(r)^2}{f(r)^2} - \frac{w(r)^3}{f(r)} = 0 \, ,
\end{equation}
where $\bar{n}$ is again an integration constant. For $d\geq 4$, there is an additional equation for the metric in the direction transverse to the condensate $h(r)$
\begin{align}
\label{B12} h''(r)&+ h'(r) \left((d-1) A'(r)+\frac{f'(r)}{f(r)}+(d-5)\frac{h'(r)}{2 h(r)}\right)- \\
\nn &-\frac{w_0^2 \ell^2}{2 N_c}\ex^{-2 A(r)}h(r)\left( w'(r)^2 + \frac{w(r)^4}{f(r)} - \frac{w(r)^2 \Phi_3(r)^2}{f(r)^2}\right) = 0 \, .
\end{align}

\subsection{Domain-wall coordinates}

In appendix \ref{AppC} are analyzed the possible IR asymptotics of the solutions to the Einstein-Yang-Mills equations. To do this analysis, it is more convenient to work with the so-called \emph{domain-wall} coordinates.

For the $(0,1)$ ansatz \eqref{B1}, the metric is written as
\be
\label{B17} \mathrm{ds}^2 = du^2 -\ex^{2A_t(u)}\intd t^2 + \ex^{2A_x(u)}( \intd x^2_1 + \dots + \intd x_{d-2}^2) + \ex^{2A_z(u)} \intd z^2 \, ,
\ee
and the equations of motion are given by
\begin{align}
\nn &A_x''(u) +\frac{1}{d-1} A_x'(u)\big(2(d-2)A_x'(u)-(d-3)(A_t'(u)+A_z'(u))\big) - \frac{2}{d-1}A_t'(u)A_z'(u)+\\
\label{B18} & + \frac{w_0^2\ell^2}{2(d-1)N_c}\ex^{-2A_t(r)-2A_z(u)}A^1_z(u)^2\Phi_3(u)^2 = 0 \, ,
\end{align}
\begin{align}
\nn &A_z''(u)+ A_z'(u) \left(A_z'(u)+ \frac{d-3}{d-1}(A_t'(u) + (d-2)A_x'(u)) \right)+\\
\nn &- \frac{d-2}{d-1}A_x'(u) \big((d-3)A_x'(u) +2A_t'(u)\big) + \\
\label{B19} &+\frac{w_0^2 \ell^2}{4 N_c}\ex^{-2 A_z(u)}\left((A^1_z)'(u)^2 - \frac{d-3}{d-1}\ex^{-2A_t(u)}A^1_z(u)^2\Phi_3(u)^2\right) = 0 \, ,
\end{align}
\begin{align}
\nn  &A_t'(u)((d-2)A_x'(u) + A_z'(u)) + \frac{d-2}{2}A_x'(u)\big((d-3)A_x'(u) + 2A_z'(u)\big) - \frac{d(d-1)}{2\ell^2} - \\
\nn &- \frac{w_0^2\ell^2}{8N_c}\ex^{-2 A_t(u)-2A_z(u)}\left(A^1_z(u)^2\Phi_3(u)^2 + \ex^{2A_t(u)} (A^1_z)'(u)^2 - \right.   \\
\label{B20} & \qquad\qquad\qquad\qquad\qquad\qquad \left.  - \ex^{2 A_z(u)}\big(\Phi_3'(u)^2 + \Phi'(u)^2\big)  \right) = 0 \, ,
\end{align}
\be
\label{B21} \Phi'(u) + \ell^{-2}\bar{n}\,\ex^{A_t(u)-(d-2)A_x(u)-A_z(u)}= 0 \, ,
\ee
\be
\label{B22} \partial_u\left(\ex^{(d-2)A_x(u)+A_z(u)-A_t(u)}\Phi_3'(u)\right) - \ex^{(d-2)A_x(u)-A_z(u)-A_t(u)}A^1_z(u)^2\Phi_3(u) = 0 \, ,
\ee
\be
\label{B23} \partial_u\left(\ex^{(d-2)A_x(u)-A_z(u)+A_t(u)}(A^1_z)'(u)\right) + \ex^{(d-2)A_x(u)-A_z(u)-A_t(u)}\Phi_3(u)^2 A^1_z(u) = 0 \, .
\ee

\vskip 2cm

For the $(1,1)$ ansatz \eqref{B9}, we write the metric as
\be
\label{B23b} \mathrm{ds}^2 = du^2 -\ex^{2A_t(u)}\intd t^2 + \ex^{2A_x(u)}( \intd x^2 + \intd y^2) + \ex^{2A_z(u)} (\intd z_1^2 + \dots + \intd z_{d-3}^2)  \, ,
\ee
and the equations of motion become
\begin{align}
\nn &A_x''(u) +\frac{1}{d-1} A_x'(u)\left(2(d-2)A_x'(u)+(d-5)\big(A_t'(u)+(d-3)A_z'(u)\big)\right) - \\
\nn &- \frac{d-3}{d-1}A_z'(u)\big(2A_t'(u) + (d-4)A_z'(u)\big)+\\
\nn & + \frac{w_0^2\ell^2}{4(d-1)N_c}\ex^{-2A_t(u)-4A_x(u)}\left((d-3) \ex^{2A_t(u)}w(u)^4 + (d-1)\ex^{2A_t(u)+2A_x(u)}w'(u)^2 - \right. \\
\label{B24} & \qquad\qquad\qquad\qquad\qquad\qquad \left.  - (d-5)\ex^{2A_x(u)}w(u)^2\Phi_3(u)^2 \right) = 0 \, ,
\end{align}
\begin{align}
\nn &A_z''(u)+ \frac{1}{d-1}A_z'(u) \left(3(d-3)A_z'(u) -(d-5)\big(A_t'(u) + 2A_x'(u)\big) \right)-\\
\nn  &- \frac{4}{d-1}A_t'(u)A_x'(u) - \frac{2}{d-1}A_x'(u)^2 + \\
\label{B25} &-\frac{w_0^2 \ell^2}{2(d-1) N_c}\ex^{-4A_x(u)}w(u)^2\left(w(u)^2 - 2 \ex^{2A_x(u)-2 A_t(u)} \Phi_3(u)^2  \right) = 0 \, ,
\end{align}
\begin{align}
\nn  &A_t'(u)(2A_x'(u) + (d-3)A_z'(u)) + A_x'(u)^2 + \frac{1}{2}(d-3)A_z'(u)\big(4A_x'(u) + (d-4)A_z'(u)\big) - \\
\nn & - \frac{d(d-1)}{2\ell^2} + \frac{w_0^2\ell^2}{8N_c}\ex^{-2 A_t(r)-4A_x(u)}\left(\ex^{2A_t(u)}w(u)^4 - 2\,\ex^{2A_x(u)}w(u)^2\Phi_3(u)^2 - \right.   \\
\label{B26} & \qquad\qquad\qquad\qquad\qquad \left. - 2\,\ex^{2A_t(r)+2A_x(u)} w'(u)^2 + \ex^{4A_x(u)}\big(\Phi_3'(u)^2 + \Phi'(u)^2\big)  \right) = 0 \, ,
\end{align}
\be
\label{B27} \Phi'(u) + \ell^{-2}\bar{n}\,\ex^{A_t(u)-2A_x(u)-(d-3)A_z(u)}= 0 \, ,
\ee
\be
\label{B28} \partial_u\left(\ex^{2A_x(u)+(d-3)A_z(u)-A_t(u)}\Phi_3'(u)\right) - 2\,\ex^{(d-3)A_z(u)-A_t(u)}w(u)^2\Phi_3(u) = 0 \, ,
\ee
\begin{align}
\nn \partial_u\left(\ex^{(d-3)A_z(u)+A_t(u)}w'(u)\right) + \ex^{(d-3)A_z(u)-A_t(u)}\Phi_3(u)^2 w(u)-\\
\label{B29} - \ex^{(d-3)A_z(u)+A_t(u) - 2A_x(u)}w(u)^3 = 0 \, .
\end{align}

\section{Physical constraints on the isospin density}

\label{CN}

We derive in this appendix the constraints on isospin density, that result from requiring the medium to be charge neutral and at $\b$-equilibrium.

We consider the medium to be composed of the strongly-coupled two-flavor quark matter described by our holographic model \eqref{Sb}, which is weakly coupled to $N_L$ species of leptons, with electric charge $-1$. Charge neutrality requires that the various particle densities are related by
\begin{equation}
	\label{CN1} \frac{1}{6} n_q + n_3 - \sum_i^{N_L} n_{L_i} = 0 \, ,
\end{equation}
where the sum is over the species of leptons that are present in the medium. Imposing also $\b$-equilibrium results in an additional condition on the chemical potentials of the particles
\begin{equation}
	\label{CN2} \mu_L\equiv \mu_{L_1} = \dots = \mu_{L_{N_L}} = -\mu_3 \, .
\end{equation}
Since thermodynamic stability requires $n_i(\mu_i)$ to be an increasing function\footnote{We also used the fact that the densities are zero at vanishing chemical potential.}, \eqref{CN2} gives the following relation between the signs of the densities
\begin{equation}
	\label{CN3} \mathrm{sign}(n_{L_1}) = \dots = \mathrm{sign}(n_{L_{N_L}}) = - \mathrm{sign}(n_3) \, .
\end{equation}
Considering positive values of $n_q$, \eqref{CN1} and \eqref{CN3} can be combined to obtain
\begin{equation}
	\label{CN4} \frac{|n_3|}{n_q} = -\frac{1}{6} \mathrm{sign}(n_3) - \frac{|Y_L|}{N_c} \, ,
\end{equation}
where we defined the total lepton fraction
\begin{equation}
	\label{CN5} Y_L \equiv \frac{\sum_i^{N_L}n_{L_i}}{n_q} \, .
\end{equation}
Equation \eqref{CN4} implies that $n_3$ must be negative, and constrained by
\begin{equation}
	\label{CN6} \frac{|n_3|}{n_q} = \frac{1}{6} - \frac{Y_L}{N_c} \, ,
\end{equation}
with $Y_L \geq 0$.

Hence, upon imposing physically relevant equilibrium conditions, the ratio of the isospin density to the quark number density is seen to have an upper bound, such that
\begin{equation}
	\label{CN7} \frac{|n_3|}{n_q} \in \left[0,\frac{1}{6}\right] \, .
\end{equation}
In the case of nuclear matter composed of protons and neutrons, the lower and upper boundaries of the interval correspond respectively to isospin symmetric matter ($Y_L = 1/2$ for $N_c = 3$), and pure neutron matter ($Y_L = 0$).

\section{Isospin density dependence of the phase transition surface}

\label{NBpt}

We derive in this appendix an analytic expression for the shape of the phase transition surface as a function of the baryon density, which is valid where the transition is second order. Knowing the line of second order phase transition at zero baryon density, this expression makes it possible to extend it to a transition surface at finite density. As discussed in the main text, the leading condensed phase is given by the $(0,1)$ solution, so we only discuss that case here.

When the phase transition is second order, the transition is signaled by the appearance of a perturbative instability on the Reissner-Nordstr\"om solution. This means that one of the quasi-normal modes for the condensing gauge fields develops a positive imaginary part. Those modes are computed from the equation of motion for the condensing gauge field $A^1_z$, which is given in \eqref{B8} and that we reproduce here for clarity
\begin{equation}
	\label{NBpt1} (A^1_z)''(r)+ (A^1_z)'(r) \left((d-3)A'(r)+\frac{f'(r)}{f(r)}-\frac{h'(r)}{2 h(r)}\right)+\frac{A^1_z(r) \Phi_3(r)^2}{f(r)^2} = 0 \, .
\end{equation}
For Reissner-Nordstr\"om, the background fields $A(r),f(r),h(r)$ and $\Phi_3(r)$ have known expressions written in appendix \ref{Sec:bgd} (in particular, $h(r)=1$).

The result that we derive originates from the observation that after inserting the RN solution, \eqref{NBpt1} depends only on two dimensionless parameters: $T/\mu_3$ and $r_H T$, with $\mu_3$ the isospin chemical potential (which is the same in all phases at the second order transition), and $r_H$ the horizon radius \eqref{Ba15}. Hence, at fixed $T/\mu_3$, the onset of the instability is controlled solely by the value of $r_H T$. From \eqref{Ba15}, this implies that the equation for the phase transition line in the $(w_0,n_3/n_q)$ plane at constant $T/\mu_3$ is of the form
\begin{equation}
	\label{NBpt5} w_0 \left(1 + \frac{n_q^2}{n_3^2}\right)^{\frac{1}{2}} = w_0^*\left(\frac{T}{\mu_3}\right) \, ,
\end{equation}
where we used that $\m_3/\m_q = n_3/n_q$ in the RN phase, and $w_0^*$ is an a priori unknown function. For a line that remains second order down to zero baryon density, $w_0^*$ is fixed by the transition line at $n_q = 0$
\begin{equation}
\label{NBpt5b} w_0^*\left(\frac{T}{\mu_3}\right) = w_0\left(\frac{T}{\mu_3},n_q=0\right) \, .
\end{equation}
This function (or rather its inverse) is shown in figure \ref{fig:PDnq4d_rqx} (rightmost line) for the case of $d=3+1$.

For a constant-$(T/\m_3)$ line that becomes first order when it reaches $n_q = 0$, the function $w_0^*$ cannot be directly inferred from the zero density phase diagram anymore. Instead, the region where the phase transition is first order needs to be computed numerically for every values of $n_q/n_3$. However, once this region is known, the function $w_0^*$ can be determined from the location of its boundary, as is clear from figure \ref{fig:PDnq4d_Tx}.

A better variable to draw the phase diagram is the chemical potential $\mu$, defined such that
\begin{equation}
	\label{NBpt6} \frac{T}{\mu}\equiv \frac{T}{\mu_3} \left(1 + \frac{n_q^2}{n_3^2}\right)^{-\frac{1}{2}}  \, .
\end{equation}
\eqref{NBpt5} can also be used as an equation for lines at constant $T/\mu$, that takes the form
\begin{equation}
	\label{NBpt7} w_0^2 \left(1 + \frac{n_q^2}{n_3^2}\right) = w_0^*\left(\frac{T}{\mu}\left(1 + \frac{n_q^2}{n_3^2}\right)^{\frac{1}{2}}\right) \, .
\end{equation}
Even though the equations of the lines are not analytic anymore, they are still completely determined in terms of the function $w_0^*$, which depends only on the data at $n_q=0$ and at the boundary of the first order transition surface.

\section{Thermodynamics around $c_3=3/2$}

\label{Thc32}

In section \ref{T0nq}, we explained that the IR expansion of the (0,1) solution at zero temperature \eqref{T0n3}, changes qualitatively at $c_3=3/2$, where $c_3$ determines the IR behavior of the isospin gauge field $\Phi_3 \sim u^{c_3}$. In this appendix, we give more details about the thermodynamic behavior of the condensed state across $c_3 = 3/2$. As in section \ref{T0nq}, we study the case of $d=3+1$, but no qualitative difference is expected for different dimensions.

\begin{figure}[h!]
\begin{center}
\includegraphics[scale=0.75]{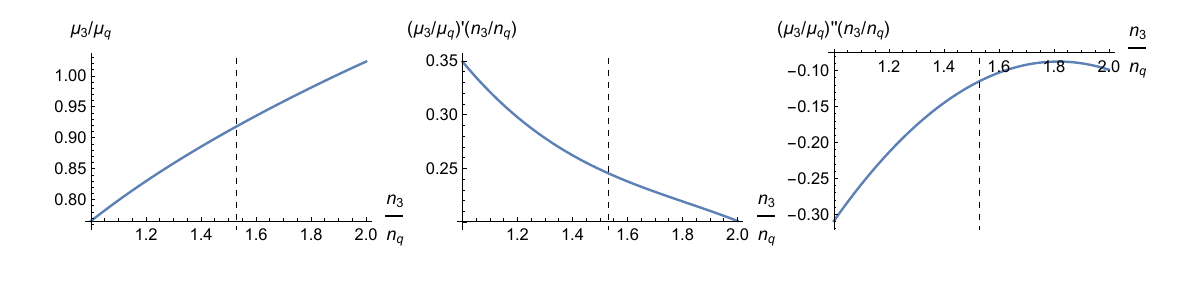}
\caption{Isospin chemical potential $\m_3$ in units of the quark number chemical potential $\m_q$, as a function of the density ratio $n_3/n_q$. We show $\m_3/m_q$ itself on the left, its first derivative with respect to $n_3/n_q$ in the middle, and the second derivative on the right. The black dashed line indicates the point where $c_3 = 3/2$. }
\label{fig:m3}
\end{center}
\end{figure}

\begin{figure}[h!]
\begin{center}
\includegraphics[scale=0.75]{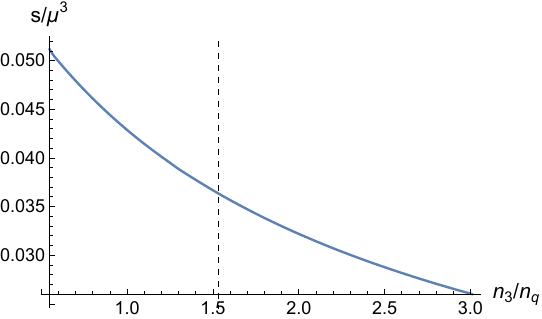}
\includegraphics[scale=0.75]{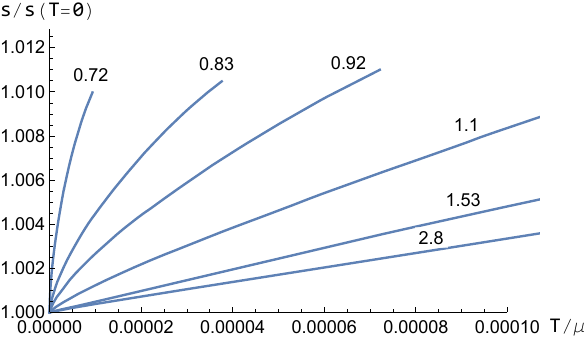}
\caption{Left : entropy density at zero temperature and $w_0 = 1$ in units of $\mu = \sqrt{\mu_3^2 + \mu_q^2}$, as a function of the ratio of isospin to quark number density $n_3/n_q$. The black dashed line indicates the point where $c_3 = 3/2$. Right : low temperature behavior of the entropy, for different values of $n_3/n_q$ indicated above each line. In particular $c_3=3/2$ is reached for $n_3/n_q \simeq 1.53$. }
\label{fig:sT}
\end{center}
\end{figure}

The type of kink observed in figure \ref{fig:D3IR} naturally raises the question whether a phase transition occurs at the corresponding point. In order to test this hypothesis, we show in figure \ref{fig:m3} the numerical result for the isospin chemical potential $\m_3$ and its low order derivatives, as a function of the parameter $n_3/n_q$ for $w_0 = 1$. These plots do not show any sign of non-analyticity when $c_3$ crosses 3/2 (indicated by the dashed line), which suggests that no finite order transition is happening there.

Even though a finite order quantum transition does not seem to occur, the low temperature thermodynamics does show a qualitative difference on the two sides of $c_3 = 3/2$. This can be observed from the behavior of the entropy density $s$. Whereas the entropy itself is smooth across the point of interest (see left figure \ref{fig:sT}), the derivative with respect to temperature goes from finite at $c_3>3/2$ to infinite at $c_3<3/2$, as indicated by the numerical results shown in right figure \ref{fig:sT}. More precisely, the low temperature scaling of the entropy is seen to change from linear above $c_3=3/2$, to a power-law smaller than 1 below. This behavior of the entropy is a further argument for something physically relevant happening at the point $c_3 = 3/2$.

\section{Analysis of the IR asymptotics: (0,1) solutions\label{G}}

In this appendix we shall give a detailed analysis of possible IR end-points  in the system of equations that describe the ground states of our problem, given in appendix \ref{AppB}. Some of them will have regular horizons while other will be extremal. We shall use the techniques pioneered in \cite{Charmousis:2010zz,Gouteraux12} to do this analysis for the (0,1) type ansatz. We shall be working in $d=3+1$. The results however generalize to other dimensions.
We do not present the case of the (1,1) ansatz as such solutions are always thermodynamically subdominant.

The metric ansatz is
\be
\label{B17a} \mathrm{ds}^2 = du^2 -\ex^{2A_t(u)}\intd t^2 + \ex^{2A_x(u)}( \intd x^2 + \intd y^2) + \ex^{2A_z(u)} \intd z^2 \, ,
\ee
We also have the three nontrivial gauge fields, $\Phi(u)$, $\Phi_3(u)$, $A^1_z(u)$. We shall call for simplicity $W(u)\equiv A^1_z(u)$ from now on.

Considering the solutions as flows in the radial direction $u$ starting at the AdS boundary, they stop in the IR when one of the slice submanifolds shrink to zero size. This can happen, generically, in two ways: either shrinking linearly so that the geometry locally is that of flat space, or exponentially so that the local geometry is that of an AdS$_p$ with $2\leq p\leq 5$.

\subsection{The time circle shrinks to zero size, $S^1_t\to 0$}

We assume that the $S^1_t$ shrinks to zero at $u=0$ by a translation in $u$. This end-point corresponds to  a non-extremal horizon. We then solve the equations asymptotically around that point, imposing regularity of the solution.

The asymptotic expansions of the various metric functions around that point are\footnote{We demand as usual that $\Phi,\Phi_3$ vanish at a regular horizon.}
\be
\ex^{A_t}={u\over \ell}\ex^{A_{t0}+{\cal O}(u^2)}\sp \ex^{A_x}=\ex^{Ax_0+{\cal O}(u^2)}\sp \ex^{A_z}=\ex^{Az_0+{\cal O}(u^2)} \, ,
\label{a1}\ee
and the three last equations (\ref{B21}), (\ref{B22}) and (\ref{B23})  become to leading order
\be
 \Phi'(u) + u\ell^{-3}\bar{n}\,\ex^{A_{t0}-(d-2)A_{x0}-A_{z0}}\simeq  0 \, ,
\ee
\be
u\pa_{u}\left({\Phi_3'\over u}\right)-\ex^{2A_{z0}}W^2\Phi_3\simeq 0 \, ,
\ee
\be
u\pa_{u}\left(u W'\right)+\ell^2 \ex^{-2A_{t0}}\Phi_3^2W\simeq 0 \, .
\ee

The solutions have regular power series expansions, with $\Phi_3$ vanishing on the horizon.
We obtain,
\be
\Phi=\Phi_0 u^2+{\cal O}(u^4)\sp \Phi_0=-{\ex^{A_{t0}- 2 A_{x0} - A_{z0}} \bar n\over 2 \ell} \, ,
\label{a2}\ee
\be
\Phi_3=\Phi_{30}u^2+{\cal O}(u^4)\sp  W=W_0+W_1u^4+{\cal O}(u^5) \, ,
\label{a3}\ee
\be
W_1=-{ \Phi_{30}^2 \ell^2\over 16 \ex^{2A_{t0}}} W_0 \, .
\label{a4}\ee
The arbitrary integration constants are $\ell,A_{t0},A_{x0},A_{z0}$, and $\Phi_{30},W_0$.

\subsection{The $z$ circle shrinks to zero, $S^1_z\to 0$}

The $\ex^{A_x}$ scale factor cannot shrink to zero, as this will give a singular solution.
Therefore, the only other scale factor that can shrink to zero is $\ex^{A_z}$.
We assume that $\ex^{A_z}\to 0$ regularly at $u=0$.

In that case $\Phi$ has a logarithmic singularity and requiring it to be absent $\Phi$ becomes constant.
Therefore such a solution exists only at zero baryon density. The scale factors are
\be
\ex^{A_t}=\ex^{A_{t0}+{\cal O}(u^2)}\sp \ex^{A_x}=\ex^{A_{x0}+{\cal O}(u^2)}\sp \ex^{A_z}={u\over \ell}\ex^{A_{z0}+A_{z1}u^2+{\cal O}(u^4)}
\label{aa1}\ee
while the solution for the gauge fields is of the form
\be
\Phi_3=\Phi_{30}\left[1+{\ell^2 W_2^2\over 16 \ex^{2A_{z0}}}u^4+\cdots\right]
\ee
\be
W=W_2u^2+W_4 u^4+\cdots
\ee
$W_{2,4}$ etc are fixed by the equations. The only constant that is not fixed apart from trivial ones is
$A_{z2}$. In this solution we also imposed the regularity condition that $W(0)=0$.

However, for this solution to make sense, $z$ must be a compact (angular) coordinate, and therefore the spatial geometry is not anymore flat infinite space, but rather an infinite cylinder.
We shall not be interested in such geometries in this paper. However, they may be relevant if the theory lives on a torus.

The above two cases exhaust all linear vanishings of scale factors. All other cases can be summarized in a generalized Lifshitz ansatz for the metric.

\subsection{Generalized Lifshitz ansatz}

We assume that as $u\to\infty$,
\be
\ex^{A_t}= \ex^{-a{u\over \ell}}+\cdots\sp    \ex^{A_x}=\ex^{-b{u\over \ell}}+\cdots  \sp \ex^{A_z}=\ex^{-c{u\over \ell}}+\cdots \, .
\label{a17}\ee
The leading curvature invariants are all constants in this case.
If the gauge fields do not back react to the metric to leading order in the IR then the metric solution is AdS$_5$ . If we want the metric to not have an IR boundary then we must have $a,b,c\geq 0$.
We shall call these metrics {\it IR-regular}.

We start by writing the leading-order Einstein equations (\ref{B18})-(\ref{B20}),
\be
b (4 b - c) - a (b + 2 c) + \zeta^2 \ex^{2 (a + c){u\over \ell}} \Phi_3^2 W^2+\cdots =0 \, ,
\label{a101}\ee
\be
-4 b (2 a + b) + 2 (a + 2 b+3c) c  -
  \zeta^2 \ex^{2 (a + c){u\over \ell}}\Phi_3^2 W^2 +
     3  \zeta^2 \ex^{2  c{u\over \ell}} (W')^2+\cdots =0
  \label{a102}\, ,   \ee
\be
-6 + b^2 + 2 b c + a (2 b + c) -{\zeta^2\over 4}\!\! \left[\ex^{2 (a + c) {u\over \ell}} \Phi_3^2 W^2\!\! -
    \ex^{2 a {u\over \ell}}((\Phi')^2+(\Phi_3')^2) \! +\!
     \ex^{2 c {u\over \ell}}(W')^2\right]+\cdots\!=0 \,,
\label{a103}\ee
with
\be
\zeta^2\equiv {w_0^2\ell^4\over 2N_c} \, .
\label{a104}\ee
It was shown in \cite{Gouteraux12} that the matter contributions in the IR limit cannot be more important than the geometry terms. Therefore, all gauge-field-related terms can be at best constants in the IR and, in particular, they cannot diverge.

The leading equation for $\Phi$ (\ref{B21}) can be integrated to give
\be
\Phi=\Phi_0+{\ell\bar n \over a-2b-c}\ex^{-(a-2b-c){u\over \ell}}+\cdots \, ,
\label{a105}\ee
or becomes linear in $u$ if $a-2b-c=0$.

In order for the solution to have $\bar n\not=0$ and still be regular we must have $a\geq 2b+c$.
This excludes $AdS_5$ ((a,b,c)=(1,1,1)),  but includes $AdS_2$,  (a,b,c)=(1,0,0).
If, on the other hand,  $\bar n=0$, then $\Phi$ is a constant and no constraint emerges for $a,b,c$.

The remaining non-abelian equations are  (\ref{B22}) and (\ref{B23}) and become

\be
\pa_u\left[\ex^{(a-2b-c){u\over \ell}}\pa_u\Phi_3\right]= \ex^{(a-2b+c){u\over \ell}}W^2\Phi_3+\cdots 
\label{a19}\ee
$$
\ar \Phi_3''+{(a-2b-c)\over \ell}\Phi_3'=\ex^{2c{u\over \ell}}W^2\Phi_3+\cdots \, ,
$$

\be
 \pa_u\left[\ex^{-(a+2b-c){u\over \ell}}\pa_u W\right]=\ex^{(a-2b+c){u\over \ell}}\Phi_3^2 W+\cdots
 \label{a20}\ee
 $$
 \ar W''-{(a+2b-c)\over \ell}W'=-\ex^{2a{u\over \ell}}\Phi_3^2 W+\cdots \, ,
 $$
 These two equations have the scaling symmetry
 \be
 \Phi_3\to \l^a\Phi_3\sp W\to \l^c W\sp u\to u-\log(\l)\ell \, ,
\label{a106}\ee
which is broken by the initial conditions.
They can be rewritten using a change of variables
\be
\Phi_3=\ex^{-a{u\over \ell}}\bar\Phi\sp W=\ex^{-c{u\over \ell}}\bar W
\label{a49}\ee
as
\be
\bar\Phi''-{(a+2b+c)\over \ell}\bar\Phi'+{a(2b+c)\over \ell^2}\bar\Phi=\bar W^2\bar \Phi+\cdots \, ,
\label{a50}\ee
\be
\bar W''-{(a+2b+c)\over \ell}\bar W'+{c(2b+a)\over \ell^2}\bar W=-\bar \Phi^2\bar W+\cdots \, .
\label{a51}\ee
The new variables are scale invariant and therefore the above equation have a genuine translational  symmetry $u\to u+\e$.

This indicates that the solution with $\bar \Phi$, $\bar W$ constants and equal to
\be
\bar W=\pm{\sqrt{a(2b+c)}\over \ell}\sp \bar \Phi=\pm{\sqrt{-c(2b+a)}\over \ell}
\label{a52}\ee
is exact (and scaling). However, according to our conventions it is not real and therefore we must abandon it. Note that the system above is also invariant under the $Z_2$ symmetry $\bar\Phi\leftrightarrow i\Bar W$,  $a\leftrightarrow c$.

We shall now try to find the solutions by checking different cases.

Another constraint is produced by the $(\Phi')^2$ term in equation (\ref{a103}). Using the solution in (\ref{a105}) we obtain that
\be
{\zeta^2\over 4}
    \ex^{2 a {u\over \ell}}(\Phi')^2={\zeta^2\ell^2\bar n^2\over 4}\ex^{2(b+c){u\over \ell}}+\cdots \, .
    \ee
Therefore for this not to diverge we must have that $b=c=0$.
This is the case we analyse further below.
Again, if $\Phi$ is constant, this constraint is void.

\subsection{AdS$_2$  $b=0,c=0,a\not=0$}

In this case the leading-order Einstein equations \eqref{a101}-\eqref{a103} take the form
\be
 \zeta^2 \ex^{2 a{u\over \ell}} \Phi_3^2 W^2+\cdots =0 \, ,
\label{a101a}\ee
\be
-
  \zeta^2 \ex^{2 a{u\over \ell}}\Phi_3^2 W^2 +
     3  \zeta^2 (W')^2+\cdots =0 \, ,
  \label{a102a}   \ee
\be
-6  -{\zeta^2\over 4} \left[\ex^{2 a {u\over \ell}} \Phi_3^2 W^2-\ell^2\bar n^2 -
    \ex^{2 a {u\over \ell}}(\Phi_3')^2  +(W')^2\right]+\cdots=0 \, ,
\label{a103a}\ee
and the U(1) gauge field is given by 
\be
\Phi=\Phi_0+{\ell\bar n \over a}\ex^{-a{u\over \ell}}+\cdots
\label{a105a}\ee

The non-abelian equations (\ref{a50})-(\ref{a51}) become
\be
\Phi_3''+{a\over \ell}\Phi_3'=W^2\Phi_3+\cdots \, ,
\label{a50a}\ee
\be
W''-{a\over \ell}W'=-\ex^{2a{u\over \ell}}\Phi_3^2 W+\cdots \, .
\label{a51a}\ee
We can again remove the explicit $u$ dependence by the redefinition
\be
\Phi_3\equiv \ex^{-a{u\over \ell}}\bar \Phi \, ,
\ee
to obtain
\be
\bar\Phi''-{a\over \ell}\bar\Phi'=W^2\bar\Phi+\cdots \, ,
\label{a50b}\ee
\be
W''-{a\over \ell}W'=-\bar\Phi^2 W+\cdots \, .
\label{a51b}\ee
The Einstein equations \eqref{a101a}-\eqref{a103a} become
\be
 \zeta^2 \bar \Phi^2 W^2+\cdots =0 \, ,
\label{a101b}\ee
\be
-
  \zeta^2 \bar \Phi^2 W^2 +
     3  \zeta^2 (W')^2+\cdots =0 \, ,
  \label{a102b}   \ee
\be
-6  -{\zeta^2\over 4} \left[ \bar \Phi^2 W^2-\ell^2\bar n^2 -
    (\bar\Phi')^2  +(W')^2\right]+\cdots=0 \, .
\label{a103b}\ee

It is clear from the equations above, that in the IR limit $u\to \infty$ we must have
\be
\bar \Phi^2 W^2\to 0\sp W'\to 0\sp \bar \Phi'^2\to {24\over \zeta^2}-\ell^2 \bar n^2 \, .
\label{g38}\ee
 These imply that $\bar \Phi$ can diverge at best linearly with $u$ and $W$ must be sublinear.
 However, there are no solutions to the equations (\ref{a50b}) and (\ref{a51b}) with $ W,\bar W$
 powerlike as $u\to\infty$.

There are two exponential  solutions of the form
\be
\bar\Phi\simeq C_3 \ex^{\kappa{u\over \ell}}\sp W\simeq C_4 \ex^{\l{u\over \ell}}
\label{a73}\ee
\begin{itemize}
\item  
\be
\l=0\sp \kappa={a-\sqrt{a^2+4\ell^2C_4^2}\over 2}<0 \, .
\label{a74}\ee

Since for this solution
$$\ex^{a{u\over \ell}}\Phi_3W=\bar \Phi W\sim \ex^{\left({a-\sqrt{a^2+4\ell^2C_4^2}\over 2}\right){u\over \ell}}\to 0\, ,$$
in the IR, (\ref{a101a}) is satisfied to leading order. Also $\ex^{a{u\over \ell}}\Phi_3',W'\to 0$ and therefore (\ref{a103a}) becomes
\be
\zeta^2\ell^2\bar n^2=24 \, ,
\ee
and  the metric is near $AdS_2$ and is supported only by the baryon gauge field, with the contributions of the non-abelian fields subleading in the IR.

\item 
\be
\kappa=0\sp \l_{\pm}={a\pm\sqrt{a^2-4\ell^{2}C_3^2}\over 2}
\label{a75}\ee
In fact this is the previous solution with the symmetry operation $\bar \Phi\leftrightarrow iW$.

However as $\l_{\pm}$ are both positive, this is not actually a solution and must be discarded.
\end{itemize}
Equations (\ref{a50b}) and (\ref{a51b}) and the conditions (\ref{g38}) imply that there are no other solutions. Note that the dependence on $a$ can be removed by scaling $a u \to u$.

\subsection{Solutions with $\Phi$=constant in the generalized Lifshitz ansatz}

Using as gauge field variables
\be
\bar W=\ex^{c{u\over \ell}}W\sp \bar \Phi=\ex^{a{u\over \ell}}\Phi_3
\ee
the Einstein equations \eqref{a101}-\eqref{a103} become
\be
b (4 b - c) - a (b + 2 c) + \zeta^2  \bar \Phi^2 \bar W^2+\cdots =0 \, ,
\label{g1}\ee
\be
-4 b (2 a + b) + 2 (a + 2 b+3c) c  -
  \zeta^2 \bar \Phi^2 \bar W^2 +
     3  \zeta^2 \left(\bar W'-{c\over \ell}\bar W\right)^2+\cdots =0 \, ,
  \label{g2}   \ee
\be
-6 + b^2 + 2 b c + a (2 b + c) -{\zeta^2\over 4} \left[\bar\Phi^2 \bar W^2 -
\left(\bar \Phi'-{a\over \ell}\bar \Phi\right)^2  +
 \left(\bar W'-{c\over \ell}\bar W\right)^2\right]+\cdots=0 \, ,
\label{g3}\ee
from which we obtain in the limit $u\to\infty$
\be
\bar \Phi^2 \bar W^2\to { a (b + 2 c)-b (4 b - c)\over \zeta^2}\geq 0 \, ,
\label{g6}\ee
\be
\left(\bar W'-{c\over \ell}\bar W\right)^2\to {3ab-(b+2c)c\over \zeta^2}\geq 0 \, ,
\label{g7}\ee
\be
\left(\bar \Phi'-{a\over \ell}\bar \Phi\right)^2\to -2{(2b+c)^2+(2b+c)a-12\over \zeta^2}\geq 0 \, ,
\label{g8}\ee
together with the gauge field equations
\be
\bar\Phi''-{(a+2b+c)\over \ell}\bar\Phi'+{a(2b+c)\over \ell^2}\bar\Phi=\bar W^2\bar \Phi+\cdots \, ,
\label{g4}\ee
\be
\bar W''-{(a+2b+c)\over \ell}\bar W'+{c(2b+a)\over \ell^2}\bar W=-\bar \Phi^2\bar W+\cdots \, .
\label{g5}\ee
Equations (\ref{g7})-(\ref{g5}) also imply that 
\be
\bar W^2\bar \Phi\to -{2b+c\over \ell}{\sqrt{-2((2b+c)^2+(2b+c)a-12)}\over \zeta}\, ,
\label{g9}\ee
\be
\bar \Phi^2\bar W\to {2b+a\over \ell}{\sqrt{3ab-(b+2c)c}\over \zeta}\, .
\label{g10}\ee
Combined with (\ref{g6}) we obtain
\be
\bar W\to {\ell\over \zeta}{a (b + 2 c)-b (4 b - c)\over (2b+a)\sqrt{3ab-(b+2c)c}}\, ,
\label{g11}\ee
\be
\bar \Phi \to -{\ell\over \zeta}{a (b + 2 c)-b (4 b - c)\over (2b+c)\sqrt{-2((2b+c)^2+(2b+c)a-12)}}\, ,
\label{g12}\ee
provided the denominators do not vanish.

Compatibility with (\ref{g4}), (\ref{g5}) requires
\begin{align}
\nn {\zeta^2\over \ell^4}&={(a (b + 2 c)-b (4 b - c))^2\over a(2b+c)(2b+a)^2(3ab-(b+2c)c)}\\
&={(a (b + 2 c)-b (4 b - c))^2\over 2c(2b+a)(2b+c)^2((2b+c)^2+(2b+c)a-12)} \, .
\label{g13}
\end{align}
However, the inequalities (\ref{g6}) and (\ref{g8}) imply that the two lines in (\ref{g13}) have opposite sign if they are non-zero. This implies that they must be zero and therefore
\be
0=a (b + 2 c)-b (4 b - c)\sim \lim_{u\to\infty}\bar \Phi^2\bar W^2 \, ,
\label{g14}\ee
from which we calculate
\be
c=b{4{b\over a}-1\over {b\over a}+2}\ar b\geq {a\over 4} \, .
\label{g15}\ee
Substituting into (\ref{g7}) we obtain
\be
1 + {b\over a} +  {b^2\over a^2} - 3 {b^3\over a^3}\geq 0\ar {b\over a}\leq 1 \, .
\label{g16}\ee
For ${1\over 4}\leq {b\over a}\leq 1$, the inequality in (\ref{g8}) is satisfied and saturated when $b=a$.

Equations (\ref{g4}) and (\ref{g5}) become
 \be
(\bar\Phi'-{a\over \ell}\bar\Phi)' -{(2b+c)\over \ell}\left(\bar\Phi'-{a\over \ell}\bar\Phi\right)\!=
(\bar\Phi'-{a\over \ell}\bar\Phi)' -{3b(a+2b)\over (b+2a)\ell}\left(\bar\Phi'-{a\over \ell}\bar\Phi\right)
\!=
\bar W^2\bar \Phi+\cdots \, ,
\label{g17}\ee
\be
\left(\bar W'-{c\over \ell}\bar W\right)'-{(a+2b)\over \ell}\left(\bar W'-{c\over \ell}\bar W\right)=-\bar \Phi^2\bar W+\cdots \, .
\label{g18}\ee

We now consider several cases
\begin{itemize}
\item Both $\bar W'-{c\over \ell}\bar W$ and $\bar\Phi'-{a\over \ell}\bar\Phi$ are non-zero constants asymptotically. Then asymptotically
\be
\bar W=C_1 \ex^{c{u\over \ell}}+C_2+\cdots\sp \bar \Phi=C_3\ex^{a{u\over \ell}}+C_4+\cdots \, ,
\label{g19}\ee
with $C_2,C_4\not= 0$. (\ref{g6}) then requires  $C_1=C_3=0$, and $C_2C_4=0$, which contradicts the assumption.
\end{itemize}

As both cannot be zero, it means that at least one of the constants to the right of
(\ref{g7}) or (\ref{g8}) must vanish.
The remaining two cases are therefore
\begin{itemize}
\item (\ref{g7}) vanishes if $b=0$ or $b=a$. In the former case  $c=0$ and this is the AdS$_2$ case studied earlier. In the latter case we obtain, $a=b=c$. In that case
\be
\left(\bar \Phi'-{a\over \ell}\bar \Phi\right)^2\to {24(1-a^2)\over \zeta^2}\geq 0
\label{g20}\ee
and we have
\be
\bar W=C_1 \ex^{a{u\over \ell}}+\cdots\sp \bar \Phi=C_2\ex^{a{u\over \ell}}+{\ell\over a}{\sqrt{24(1-a^2)}\over \zeta}+\cdots \, .
\ee
Compatibility with (\ref{g6}) requires that $C_2=0$ and $a=1$\footnote{An attempt to make $C_1$ vanish instead does not work because of the minus sign in the right-hand side of (\ref{g22}).}. Therefore $a=b=c=1$ and we obtain the AdS$_5$ IR geometry.

In this case equations (\ref{g4}), (\ref{g5}) become
\be
\bar \Phi''-{4\over \ell}\bar \Phi'+{3\over \ell^2}\bar \Phi=\bar W^2\bar \Phi \, ,
\label{g21}\ee
\be
\bar W''-{4\over \ell}\bar W'+{3\over \ell^2}\bar W=-\bar \Phi^2\bar W \, .
\label{g22}\ee

The non-trivial equation to leading order is (\ref{g21}) and has a double exponential solution
\be
\bar \Phi\sim \ex^{-|C_1|\ex^{u\over \ell}}+\cdots \, .
\label{g23}\ee
This solution is discussed in more detail in section \ref{Sec:T0}.

\item (\ref{g8}) vanishes if\footnote{There are other roots, but they make (\ref{g7}) negative or complex.}
\be
b=-{7 a\over 24} + {\sqrt{48 + a^2}\over 24} + {\sqrt{
 24 - 23 a^2 + {1968 a\over \sqrt{48 + a^2}}+ {41 a^3\over \sqrt{48 + a^2}}}\over
 12 \sqrt{2}}\sp 1\leq a\leq 4 \, .
 \label{g24}\ee
 We also have
 \be
 c={1\over 12} \left(a + 5 \sqrt{48 + a^2} - \sqrt{48 - 46 a^2 + 82 a \sqrt{48 + a^2}}\right) \, .
 \ee
Since $a\in[1,4]$, $c\in [1,0]$.

We obtain
\be
\left(\bar W'-{c\over \ell}\bar W\right)^2\to w^2 \, ,
\label{g25}\ee
with
\be
w^2=
{-144 -  \left(a + \sqrt{48 + a^2}\right) \left(6 a - \sqrt{
    48 - 46 a^2 + 82 a \sqrt{48 + a^2}}\right)\over  8\zeta^2}\geq 0 \, ,
\label{g26}\ee
so that
\be
\bar \Phi=C_1 \ex^{a{u\over \ell}}+\cdots\sp \bar W=C_2\ex^{c{u\over \ell}}+{\ell\over c}w+\cdots
\label{g27}\ee
Compatibility with (\ref{g6}) again requires

(a) $C_2=0$ and $w=0$. In that case $a=b=c=1$ and we recover again AdS$_5$.
However, there is no compatible sufficiently vanishing solution for $\bar W$.

(b) $C_1=0$. In this case there is a consistent solution for $\bar \Phi$
\be
\bar \Phi\sim \ex^{-{|C_2|\over c}\ex^{c{u\over \ell}}}\,.
\ee
Substituting this solution in (\ref{g18}), we obtain that $w=0$, which implies $a=1$. These are again the AdS$_5$ asymptotics found earlier.
\end{itemize}

This concludes our analysis on the possible IR asymptotics of our Einstein-YM equations in the (0,1) ansatz.

\label{AppC}

\section{Perturbative stability near the extremal Reissner-Nordstr\"om horizon}

\label{appH}

In this appendix, we analyze the stability of the AdS$_2$ geometry near the horizon of the extremal black-hole, with respect to condensation of the gauge field.

The extremal RN solution corresponds to the zero temperature limit of the solution presented in appendix \ref{Sec:bgd}. The near-horizon limit is obtained by considering
\be
\label{D1} r = r_H \left(1- \e \frac{r_H\ell_2}{\ell^2}\frac{\ell_2}{\zeta}\right) \sp t = \\ex^{-1} \tau \sp \ell_2 \equiv \frac{\ell}{\sqrt{d(d-1)}} \sp \e \ll 1 \, ,
\ee
such that, at leading order in $\e$, the metric is that of AdS$_2\times \mathbb{R}^{d-1}$
\begin{equation}
\label{D1b} \intd s^2 = \left(\left(\frac{\ell_2}{\zeta}\right)^2 \left(-\intd\tau^2 + \intd\z^2 \right) + \left(\frac{\ell}{r_H}\right)^2\intd\vec{x}^2 \right) \Big[1 + \mathcal{O}(\e)\Big] \, .
\end{equation}
$\zeta$ is the AdS$_2$ radial coordinate, with the boundary located at $\zeta = 0$, and $\tau$ is the time coordinate.

An instability of AdS$_2$ will arise when one of the gauge field perturbations on the RN background exhibits a growing mode in the near-horizon region. By invariance of RN under spatial rotations and chiral rotations in the (1,2) plane, it is enough to study a restricted set of four independent perturbations, that can be chosen to be $\d A^1_t,\d A^1_z,\d A^3_t$ and $\d A^3_z$. The equations of motion that they obey are given by the linearized Yang-Mills equations \eqref{PD2}. For $\d A^3_t$ and $\d A^1_z$, those are the linearization of \eqref{B7} and \eqref{B8},
\begin{equation}
\label{D1c} \partial^2_r \d A^3_t(r) - \frac{d-3}{r} \partial_r \d A^3_t(r) = 0 \, ,
\end{equation}
\begin{equation}
\label{D1d} \partial^2_r \d A^1_z(r) + \left(\frac{f'(r)}{f(r)} - \frac{d-3}{r} \right)\partial_r \d A^1_z(r) + \frac{\Phi_3(r)^2}{f(r)^2}\d A^1_z(r)  = 0 \, ,
\end{equation}
where $\Phi_3(r)$ and $f(r)$ are the fields of the background, given in \eqref{Ba10} and \eqref{Ba16}. As for $\d A^1_t$ and $\d A^3_z$, it is not difficult to show that they obey
\begin{equation}
\label{D1e} \partial^2_r \d A^1_t(r) -\frac{d-3}{r} \partial_r \d A^1_t(r) = 0 \, .
\end{equation}
\begin{equation}
\label{D1f} \partial^2_r \d A^3_z(r) + \left(\frac{f'(r)}{f(r)} -\frac{d-3}{r}\right) \partial_r \d A^3_z(r) = 0 \, .
\end{equation}

The near-horizon limit of those equations is obtained  by applying the change of coordinate \eqref{D1}, and keeping only the leading order in $\e$. This results in
\be
\label{D2a} (\d A^a_t)''(\zeta) + \frac{2}{\zeta}(\d A^a_t)'(\zeta) = 0 \, , \quad a = 1,3 \, ,
\ee
\be
\label{D2b} (\d A^3_z)''(\zeta) = 0 \, ,
\ee
\be
\label{D2c} (\d A^1_z)''(\zeta) + \left(  \frac{(d-2)\m_3 r_H}{d(d-1)\zeta}\right)^2 \d A^1_z(\zeta) = 0 \, ,
\ee
which are respectively the equations in AdS$_2$ for two massless gauge fields, a massless scalar field, and a massive scalar field with mass
\be
\label{D3} m^2 \ell_2^2 = -\left(\frac{(d-2)\m_3 r_H}{d(d-1)}\right)^2 \, .
\ee

The AdS$_2$ instability sets in when the mass squared of a mode gets below the AdS$_2$ Breitenlohner-Freedman (BF) bound
\be
\label{D4} m_{\text{BF}}^2 \ell_2^2 = - \frac{1}{4} \, ,
\ee
which can only happen for $\d A^1_z$. Using \eqref{D3} and the known expression of the Reissner-Nordstr\"om horizon radius \eqref{Ba14}, the instability is found to happen for $w_0$ below a critical value
\be
\label{D5} w_{0,c} = 4\sqrt{\frac{2N_c}{d(d-1)N_f(1+(n_q/n_3)^2)}} \, .
\ee
If no transition happens before that, it is expected that an infinite order Berezinsky-Kosterlitz-Thouless (BKT)-like instability sets in at $w_{0,c}$ \cite{Jensen10,Iqbal10,V1b,V6}.

\newpage

\end{document}